\documentclass[aps,twocolumn,prd,preprintnumbers,superscriptaddress,nofootinbib,floatfix,balancelastpage]{revtex4-1}

\usepackage{amsmath,amsfonts,amssymb}
\usepackage[T1]{fontenc}
\usepackage[utf8]{inputenc}
\usepackage[british]{babel}
\usepackage{lmodern}
\usepackage{microtype}
\usepackage{tikz}
\usepackage{slashed}
\usepackage{placeins}
\usepackage[compat=1.1.0]{tikz-feynhand}
\usepackage{comment}
\usepackage{xfrac}
\usepackage{tikz}
\usepackage{color}
\usepackage{xcolor,colortbl}

\usepackage{bm}
\usepackage{url}
\usepackage[colorlinks=true,urlcolor=blue,linkcolor=blue,citecolor=blue,hypertexnames]{hyperref}

\definecolor{Gray}{gray}{0.85}
\definecolor{LightGreen}{rgb}{0.88, 1, 0.88}
\definecolor{Blue}{rgb}{0,1,1}
\definecolor{Lime}{rgb}{0,1,0}
\definecolor{LightCyan}{rgb}{0.88,1,1}
\definecolor{LightRed}{rgb}{1, 0.85, 0.85}
\definecolor{Red}{rgb}{1, 0, 0}
\definecolor{LightYellow}{rgb}{1, 1, 0.85}
\definecolor{Yellow}{rgb}{1,1,0.05}
\definecolor{LightBlue}{rgb}{0.87, 0.94, 1}
\definecolor{white}{gray}{1}
\definecolor{black}{gray}{0}

\definecolor{lightgray}{gray}{0.91}
\definecolor{LightGray}{gray}{0.93}

\newcolumntype{?}{!{\vrule width 1pt}}
\newcolumntype{`}{!{\vrule width 1.5pt}}

\newcommand{\eq}[1]{\eqref{eq:#1}}
\newcommand{\fig}[1]{Fig.~\ref{fig:#1}}
\newcommand{\tab}[1]{Tab.~\ref{tab:#1}}
\newcommand{\Sec}[1]{Sec.~\ref{sec:#1}}
\newcommand{\App}[1]{App.~\ref{sec:#1}}
\newcommand{\tr}{\mathrm{tr}\!}

\definecolor{MathBlue}{rgb}{0.368417, 0.506779, 0.709798}
\definecolor{MathYellow}{rgb}{0.880722, 0.611041, 0.142051}
\definecolor{MathGreen}{rgb}{0.560181, 0.691569, 0.194885}
\definecolor{MathDarkGreen}{rgb}{0.571589, 0.586483, 0.}
\definecolor{MathRed}{rgb}{0.922526, 0.385626, 0.209179}
\definecolor{MathViolet}{rgb}{0.528488, 0.470624, 0.701351}
\definecolor{MathPurple}{rgb}{0.647624, 0.37816, 0.614037}
\definecolor{MathBrown}{rgb}{0.772079, 0.431554, 0.102387}

\hypersetup{
	colorlinks,
	linkcolor={blue!75!black},
	citecolor={blue!75!black},
	urlcolor={blue!75!black}
}

\usepackage{array,mathtools,amssymb,booktabs}
\newcolumntype{C}{>{$}c<{$}}
\AtBeginDocument{
\heavyrulewidth=.16em
\lightrulewidth=.1em
\cmidrulewidth=.03em
\belowrulesep=.4ex
\belowbottomsep=0pt
\aboverulesep=.4ex
\abovetopsep=0pt
\cmidrulesep=\doublerulesep
\cmidrulekern=.5em
\defaultaddspace=.5em
}

\makeatletter
    \def\CT@@do@color{%
      \global\let\CT@do@color\relax
            \@tempdima\wd\z@
            \advance\@tempdima\@tempdimb
            \advance\@tempdima\@tempdimc
    \advance\@tempdimb\tabcolsep
    \advance\@tempdimc\tabcolsep
    \advance\@tempdima2\tabcolsep
            \kern-\@tempdimb
            \leaders\vrule
                    \hskip\@tempdima\@plus  1fill
            \kern-\@tempdimc
            \hskip-\wd\z@ \@plus -1fill }
    \makeatother

\begin{document}

\title{Asymptotic Safety Guaranteed at Four Loop}%
\preprint{DO-TH 22/13}

\author{Daniel F. Litim}
\author{Nahzaan Riyaz}
\affiliation{
Department of Physics and Astronomy, University of Sussex, Brighton, BN1 9QH, United Kingdom
}
\author{Emmanuel Stamou}
\affiliation{
	Fakultät für Physik, TU Dortmund, Otto-Hahn-Str.~4, D-44221 Dortmund, Germany
}

\author{Tom Steudtner} 
\affiliation{
	Fakultät für Physik, TU Dortmund, Otto-Hahn-Str.~4, D-44221 Dortmund, Germany
}
\affiliation{Department of Physics, University of Cincinnati, Cincinnati, OH 45221, USA}
\begin{abstract}

We investigate a family of four-dimensional quantum field theories  
with weakly interacting ultraviolet   fixed  points up to four loop order in perturbation theory.
Key new ingredients are  the three loop  gauge contributions to quartic scalar beta functions, which we compute in  the $\overline{\text{MS}}$
scheme for a template $SU(N_c)$ gauge theory  coupled to $N_f$  fundamental fermions and elementary scalars.
We then determine  fixed point couplings,  field and mass anomalous dimensions, and universal scaling exponents up to the first three non-trivial orders in a small Veneziano parameter.
 The phase diagram and  UV-IR connecting trajectories are found and contrasted with asymptotic freedom.
 Further, the size of the conformal window, unitarity, and mechanisms leading to the loss of conformality are investigated. 
 Our results provide blueprints for concrete 4d non-supersymmetric  conformal field theories with standard model-like field content and invite further model building.
\end{abstract}

\maketitle

\tableofcontents

\section{Introduction}

Ultraviolet (UV) fixed points play a central role for the fundamental definition of quantum field theory (QFT). 
They ensure that theories are UV-complete, meaning  well-defined and predictive up to highest energies.
This is quite different 
 from effective field theories  that  tend to break down above a certain energy. 
Moreover, and much like critical points in systems with continuous phase transitions,  fixed points in particle physics  also relate to an underlying  conformal field theory (CFT).  
The existence of free UV fixed points, known as asymptotic freedom, has been established  long ago~\cite{Gross:1973id,Politzer:1973fx}.
The more recent discovery 
that high-energy fixed points can  also be interacting   \cite{Litim:2014uca,Bond:2016dvk,Bond:2018oco,Bond:2017lnq,Bond:2017suy,Bond:2017tbw,Bond:2017sem,Buyukbese:2017ehm,Bond:2019npq,Litim:2020jvl,Bond:2021tgu,Bond:2022xvr}, known as asymptotic safety~\cite{Bailin:1974bq,Weinberg:1980gg}, has  opened up new territory to look for UV-complete extensions of the Standard Model, 
 and for genuinely new phenomena beyond the paradigms of asymptotic freedom or effective field theory 
\cite{Litim:2015iea,
Sannino:2014lxa,
Nielsen:2015una,
Rischke:2015mea,
Codello:2016muj,
Bond:2017wut,
Dondi:2017civ,
Kowalska:2017fzw,
Abel:2017ujy,
Christiansen:2017qca,
Sannino:2018suq,
Barducci:2018ysr,
Hiller:2019mou,
Hiller:2019tvg,
Hiller:2020fbu,
Bissmann:2020lge,
Bause:2021prv,
Hiller:2022hgt,
Hiller:2022rla,
Hiller:2023bdb}.

\begin{figure}
    \centering
\vskip-.5cm    \includegraphics[width=.4\textwidth]{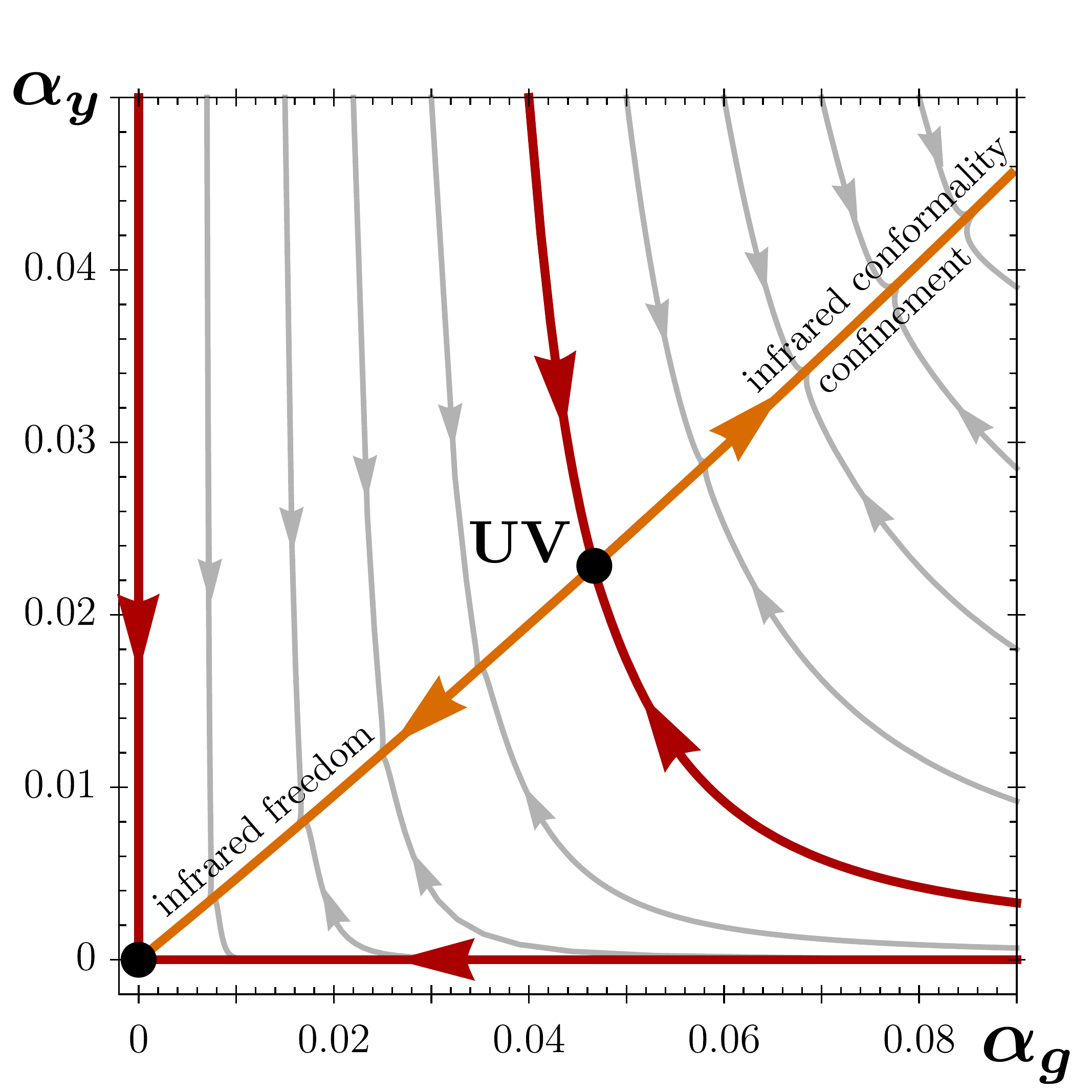}
    \caption{Phase diagram of an asymptotically safe theory in the gauge-Yukawa plane of couplings $(\alpha_g,\alpha_y)$ at four loop. Shown are the 
    interacting  UV and the free IR  fixed points (black dots), and sample trajectories with arrows pointing to the IR. Two asymptotically safe trajectories are running out of the UV fixed point  (orange) lead either to infrared freedom at weak coupling, or to confinement or conformality at strong coupling.
 }
    \label{fig:PD}
\end{figure}

A role model for an UV-complete particle theory with a weakly interacting fixed point  
is given by $N_f$   fermions coupled to $SU(N_c)$ gauge fields and elementary scalars
through gauge and Yukawa interactions   
\cite{Litim:2014uca}. 
Crucially, in the regime where asymptotic freedom is absent, quantum fluctuations ensure that the  growth of the gauge coupling  
is countered by Yukawa couplings, leading to an interacting fixed point in the UV (see  \fig{PD}). 
In the large-$N$  limit, the  fixed point is under strict perturbative control,  and  specifics of the theory can be extracted systematically in perturbation theory by using  $\epsilon=N_f/N_c-11/2$ as a small control parameter. 
Thus far, critical couplings, universal  exponents, and the size of the conformal window have been determined  up to the second non-trivial order in $\epsilon$, including finite $N$ corrections   \cite{Litim:2014uca,Litim:2015iea,Bond:2017tbw,Bond:2021tgu}.

In this paper, we extend the study of the UV critical theory to the complete third order  in $\epsilon$. 
The rationale for this is that while the  fixed point occurs for the first time at the leading order in $\epsilon$ \cite{Litim:2014uca}, a bound on   the UV conformal window $\epsilon<\epsilon_{\rm max}$ arises for the first time at the complete second order  in $\epsilon$ \cite{Bond:2017tbw,Bond:2021tgu}. 
Thereby, it has also been noted that $\epsilon_{\rm max}$ is numerically small, suggesting that the entire UV conformal window could be within the perturbative domain.\footnote{This state of affairs should be compared with the IR conformal window of $SU(N_c)$ gauge theories coupled to $N_f$ fundamental fermions, where the theory becomes strongly coupled; see  for instance \cite{DelDebbio:2010zz,DiPietro:2020jne,Kluth:2022wgh} and references therein.}
The validation of this picture warrants a study up to the third non-trivial order in $\epsilon$. 
To achieve this goal,  the  four-loop gauge,  three-loop Yukawa and quartic $\beta$ functions, and three-loop  anomalous dimensions  are required as input. Some of these can be extracted from generic expressions for $\beta$ functions of gauge-Yukawa theories~\cite{Bednyakov:2021qxa,Davies:2021mnc}. The missing pieces, however, are the three-loop contributions to scalar $\beta$ functions containing gauge interactions, which  we compute  using standard techniques in the $\overline{\text{MS}}$ scheme, and which is one of the central results of this work. In addition, we provide fixed point couplings and  conformal data of the UV critical theory up to  cubic order in $\epsilon$, and look into the loss of conformality, the range of perturbativity, and  UV-IR connecting trajectories in comparison with asymptotic freedom.

The paper is organised as follows.  \Sec{Determination} recalls the basics of asymptotically safe particle theories, introduces our  model, and explains the underlying systematics.
In \Sec{computation}, we detail the computation of $\beta$ functions. In \Sec{betafunction}, we present our results which include $\beta$ functions, fixed points, anomalous dimensions and scaling exponents,  bounds on the conformal window, and aspects of the phase diagram.
We conclude in  \Sec{conclusion}, and defer additional material to three appendices.

\section{Asymptotically Safe Gauge Theory}\label{sec:Determination}
In this section, we recall the basic model and its quantum critical points, and the systematics of the underlying perturbative and conformal expansions.
\subsection{Model}\label{sec:setup}
We consider a four-dimensional, renormalisable QFT with $SU(N_c)$ gauge group and an unbroken $U(N_f)_L \times U(N_f)_R$ global flavour symmetry. In analogy to massless QCD, the theory features $N_f$ coloured, quark-like fermion as well as a complex but uncharged meson-matrix scalar, as listed in \tab{matter}. 
\begin{table}[b]
\centering
 	\aboverulesep = 0mm
	\belowrulesep = 0mm
	\addtolength{\tabcolsep}{1pt}
	\setlength{\extrarowheight}{1pt}
    \begin{tabular}{`lcccc`}
    \toprule   
         \rowcolor{LightBlue}
         Field && $SU(N_c)$ & $U_L(N_f)$ & $U_R(N_f)$ \\
         \hline
         Fermion & $\ \psi_L$ & $N_c$ & $N_f$ & $1$ \\ \rowcolor{LightGray} 
         & $\ \psi_R$ & $N_c$ & $1$ & $N_f$ \\[.1em]  
         \  Scalar & $\ \phi$ & $1$ & $N_f$ & $\overline{N_f}$\\[.1em]
\toprule
    \end{tabular}
    \caption{Field content and representations under gauge and global symmetry.}
    \label{tab:matter}
\end{table}
The corresponding Lagrangian
\begin{equation}\label{eq:lag}
    \begin{aligned}
       \mathcal{L} &=  - \tfrac14\,F^{A\mu\nu} F^A_{\phantom{A}\mu\nu} + \mathcal{L}_\text{gf} + \mathcal{L}_\text{gh} \\
       &\phantom{= }  + \tr\left[\overline{\psi} i \slashed{D}\psi\right] - y\,\tr\left[\overline{\psi} \left( \phi \,\mathcal{P}_R + \phi^\dagger \,\mathcal{P}_L\right)\psi\right]  \\
       &\phantom{= } + \tr\left[\partial^\mu \phi^\dagger \partial_\mu \phi\right]  - m^2\,\tr\left[\phi^\dagger  \phi\right] \\
       &\phantom{= } - u\,\tr\left[\phi^\dagger  \phi\phi^\dagger  \phi\right] - v\,\tr\left[\phi^\dagger  \phi\right] \,\tr\left[\phi^\dagger  \phi\right]
    \end{aligned}
\end{equation}
consists of a gauge sector with field strength tensor $F^A_{\phantom{A}\mu\nu}$, the usual gauge fixing and ghost terms $\mathcal{L}_\text{gf}$ and $\mathcal{L}_\text{gh}$ and the coupling to the fermions via the covariant derivative $D_\mu$ and the
gauge coupling $g$.
Traces in~\eq{lag} run over both flavour and gauge indices.
Cross-talk between the scalar and gauge sector is  mediated via  the real Yukawa   coupling $y$. This interaction is manifestly chiral due to the projectors $\mathcal{P}_{R,L} = \tfrac12(1\pm \gamma_5)$. 
In the scalar sector, we observe
real-valued single-trace $(u)$ and double-trace quartic couplings $(v)$.
The scalar mass term  in~\eq{lag} is compatible with the global symmetry of the model.
Below, we are mostly interested in the massless limit.

\subsection{Veneziano Limit}
In this work, we are interested in the  planar (Veneziano) limit~\cite{Veneziano:1976wm},
where field multiplicities  $N_f$ and $N_c$ are large and interactions are parametrically weak.
The virtue of the Venziano limit is that it offers rigorous perturbative control, allowing systematic 
expansions in a small parameter.
To prepare for the Veneziano limit, we introduce rescaled couplings~\cite{tHooft:1973alw}
\begin{equation}\label{eq:tHooft-couplings}
    \alpha_x= \frac{N_c\,x^2}{(4 \pi)^2}\,, 
    \qquad \alpha_u = \frac{N_f\,u}{(4 \pi)^2}\,, 
    \qquad \alpha_v = \frac{N_f^2\,v}{(4 \pi)^2}\,,
\end{equation}
where $x=g,y$. Notice that the gauge, Yukawa, and single-trace scalar couplings scale linearly, while the double-trace scalar  couplings scales quadratically with matter field multiplicity. In the Veneziano limit, any explicit dependence on $(N_c, N_f) $ drops out after the rescaling \eqref{eq:tHooft-couplings}, and leaves us with a dependence on $\epsilon$,
\begin{equation}\label{eq:eps}
\epsilon \equiv  \frac{N_f}{N_c} - \frac{11}{2}\,.
\end{equation}
Moreover, the parameter  \eqref{eq:eps}  becomes  continuous in this limit, taking values in the entire range $\epsilon\in [-\tfrac{11}{2},\,\infty )$.
We are particularly interested in the  regime   
\begin{equation}
|\epsilon| \ll 1
\end{equation}
where it serves as a small control parameter for perturbativity.
The virtue of the parameter \eqref{eq:eps} is that it is
proportional to the one-loop coefficient of the gauge $\beta$
function, which is at the root for perturbatively controlled fixed points in any 4d quantum field theory \cite{Bond:2016dvk,Bond:2018oco}.

This last point can be illustrated, exemplarily, by expanding a gauge $\beta$ function to second loop order, 
\begin{equation}
    \beta_{g}\big|_\text{null} = \tfrac43 \epsilon\, \alpha_g^2 + C\,\alpha_g^3 + \mathcal{O}(\epsilon\,\alpha_g^3,\alpha_g^4)\,.
\end{equation}
If other couplings $\alpha_i$ are present, we project them onto their nullclines ($\beta_i=0$). The coefficient $C$, generically of order unity,  relates to the gauge two loop coefficient, possibly modified through Yukawa interactions by the nullcline projection  \cite{Bond:2016dvk}. 
Consequently,
 a non-trivial fixed point arises from a cancellation between the parametrically suppressed one-loop term and the two-loop term,
\begin{equation} \label{eq:ag-expand}
    \alpha_g^* =  - \frac{4 \,\epsilon}{3 C}  +  \mathcal{O}(\epsilon^2)\,,
\end{equation}
 leading to a power series in the control parameter $\epsilon$, with higher order loop terms leading to subleading corrections in $\epsilon$.\footnote{Physicality of the fixed point requires that $\epsilon\cdot C<0$.}
If other couplings are present, the nullcline conditions dictate that their fixed points are  $\alpha_i^*\propto \alpha_g^*$. We conclude that strict perturbativity of  fixed points in non-abelian gauge theories can always be guaranteed
for sufficiently small $\epsilon \to 0$  \cite{Bond:2016dvk,Bond:2018oco}.  For examples of gauge theories where interacting UV fixed points exist non-perturbatively, including away from  a Veneziano limit and  at large $\epsilon$, we refer to \cite{Bond:2022xvr}.

\subsection{Systematics} 
A key feature of non-abelian gauge theories coupled to matter 
is that fixed point couplings $\alpha_i^*$ \eqref{eq:tHooft-couplings} can be systematically expanded as a power series in the small parameter $\epsilon$ \cite{Bond:2017tbw}. For our setting, this implies the ``conformal expansion'' in powers of $\epsilon$,
\begin{equation}\label{eq:FP-exp}
	\alpha_i^* = \sum_{n=1}^\infty  \alpha_{i}^{(n)}\, \epsilon^n \,,\quad (i=g,y,u,v)\,.
\end{equation}
The expansion coefficients $ \alpha_{i}^{(n)}$ are determined using perturbation theory.
In order to obtain all fixed point couplings \eqref{eq:FP-exp} accurately up to and including the order $\epsilon^n$, 
the perturbative loop expansion must be performed up to the loop order $n+1$ in the gauge, and up to order $n$ in the Yukawa and quartic $\beta$ functions, to which we refer as the   \texttt{(n+1)nn} approximation \cite{Bond:2017tbw}.\footnote{For want of terminology, we denote settings which retain the gauge/Yukawa/quartic $\beta$ functions up to  the $l/m/n$ loop order as
the ``\texttt{lmn} approximation.''} Ultimately, the reason why the systematics of the perturbatively controlled expansion requires one more loop order  in the gauge sector is that  the one-loop gauge coefficient is parametrically as large as the gauge two-loop coefficient. 
This result establishes a link between the perturbative loop expansion and the conformal expansion in $\epsilon$. The leading order 
$\epsilon^0$ (LO) relates to the loop order  \texttt{100}, where the running gauge coupling is parametrically slowed down but a fixed point cannot (yet) arise. The next-to-leading order $\epsilon^1$ (NLO), corresponding  to \texttt{211}, offers the first  non-trivial order where a fixed point  materialises \cite{Litim:2014uca}, and the next-to-next-to-leading order $\epsilon^2$  (2NLO), 
corresponding to  \texttt{322}, is the first non-trivial order where bounds on the conformal window arise  \cite{Bond:2017tbw,Bond:2021tgu}. 
In this work, we provide the  order $\epsilon^3$ (3NLO) corresponding to the \texttt{433} approximation.

\subsection{Fixed Points}
\begin{figure}
    \centering
\begin{tikzpicture}
  \shade[bottom color=MathViolet, top color=white] (-5, 0) rectangle (-1.9, +.6);
  \shade[top color=MathPurple, bottom color=white] (-3.4, 0) rectangle (-1.9, -.6);
  \shade[bottom color=MathGreen, top color=white] (-1.9, 0) rectangle (1.2, +.6);
  \shade[top color=MathDarkGreen, bottom color=white] (-1.9, 0) rectangle (1.2, -.6);
  \shade[top color=MathRed, bottom color=white]  (-5,0) rectangle (-3.4, -.6);
  \shade[bottom color=MathYellow, top color=white]  (1.2,0) rectangle (2.6, .6);
  \shade[bottom color=white, top color=MathBrown]  (1.2,0) rectangle (2.6, -.6);
  \draw (-3.4, +.3) node  {\scriptsize \textsf{Asymptotic Freedom}};   
  \draw (-.4, +.3) node  {\scriptsize \textsf{Asymptotic Safety}};
  \draw (2., +.3) node  {\scriptsize \textsf{\begin{tabular}{c} Effective \\ Theory \end{tabular}}};
  \draw (2., -.35) node  {\scriptsize \textsf{\begin{tabular}{c} IR \\ Freedom \end{tabular}}};
  \draw (-0.4, -.35) node  {\scriptsize \textsf{ \begin{tabular}{c}  IR Freedom or \\ Confinement/Conformality \end{tabular}}};

  \draw (-2.65, -.32) node  {\scriptsize \textsf{Banks-Zaks}};
  \draw (-4.2, -.32) node  {\scriptsize \textsf{Confinement}};

  \draw[thick]  (2.6, 0) -- (-5, 0) -- (-5, +.6) -- (-5, -.6);
  \draw[thick]  (-1.9, +.6) -- (-1.9, -.6);
  \draw[thick]  (-3.4, 0) -- (-3.4, -.6);
  \draw[thick]  (1.2, -.6) -- (1.2, +.6);
  \fill[fill=black] (2.8,0) -- (2.6,.1) -- (2.6,-.1);
  
  \draw (3, 0) node  {\Large $\mathbb{\epsilon}$};
  \draw (-5.25, +.3) node  {\scriptsize  \textbf{\textsf{UV}}};   
  \draw (-5.25, -.35) node  {\scriptsize   \textbf{\textsf{IR}}};   
  \draw (-5, -1.) node  {\centering $-\tfrac{11}2$};
  \draw (-1.9, -1.) node  {\centering $0$};
  \draw (-3.4, -1.) node  {\centering $\epsilon_\text{min}$};
  \draw (1.2, -1.) node  {\centering $\epsilon_\text{max}$};

  \end{tikzpicture}
    \caption{Main characteristics of the theory \eqref{eq:lag}  as a function of the Veneziano parameter $\epsilon$. In the UV, we indicate whether the theory is asymptotically free,  safe, or UV-incomplete and described by an effective field theory. In the IR, we indicate whether the theory achieves confinement, IR freedom, or an interacting conformal fixed point.}
    \label{fig:Schematic}
\end{figure}
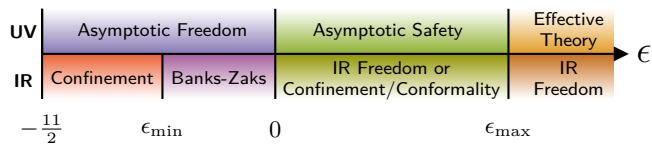

We  briefly recall  the weakly interacting fixed points of the theory \eq{lag}.
For  $ \epsilon < 0$, the theory is asymptotically free~\cite{Gross:1973id,Politzer:1973fx}
and one  finds the seminal Caswell-Banks-Zaks IR fixed point~\cite{Caswell:1974gg,Banks:1981nn} with $\alpha_g^* > 0$ and  $\alpha_{y,u,v}^* = 0$.
The IR fixed point is known to exist within a conformal window $\epsilon_\text{min} < \epsilon < 0$, analogous to the conformal window in QCD with extra fermions. The upper end is determined by the loss of asymptotic freedom. The fixed point becomes strongly coupled at the lower end $\epsilon=\epsilon_\text{min}$. The exact value for $\epsilon_\text{min}> -\tfrac{11}2$, however, is not established with high accuracy, 
see  for instance \cite{DiPietro:2020jne,Kluth:2022wgh} and references therein.
Also, in the regime  with asymptotic freedom, the theory  does not exhibit a perturbatively controlled 
fixed point with  non-trivial Yukawa interactions $\alpha_y^* \neq 0$~\cite{Bond:2016dvk}. These main characteristics  are  illustrated in \fig{Schematic}.

For $\epsilon > 0$, on the other hand, asymptotic freedom is absent. 
Then, a UV completion  requires the appearance of an interacting UV fixed point. 
Most importantly, such a phenomenon 
necessitates a delicate interplay of non-abelian gauge, Yukawa and scalar interactions, and cannot arise from gauge interactions alone  \cite{Bond:2016dvk,Bond:2018oco}.
It then gives rise to a fully interacting UV fixed point $(\alpha_{g,y,u,v}^* \neq 0)$~\cite{Litim:2014uca,Litim:2015iea} and a conformal  window $0 < \epsilon < \epsilon_\text{max}$. 

This UV fixed point and its renormalisation group (RG) flow in the $(\alpha_g,\,\alpha_y)$ plane  is shown in \fig{PD}.
In the vicinity the UV fixed point, the RG flow is power-law rather than logarithmic, with respect to the renormalisation scale~$\mu$
\begin{equation}\label{eq:power-law-running}
   \alpha_i = \alpha_i^* + \sum_jc_{i,j} \left(\frac{\mu}{\mu_0}\right)^{\vartheta_j}\,,
\end{equation}
characterised by universal scaling exponents $\vartheta$. 
The sign of the critical exponents $\vartheta_i$ in \eq{power-law-running} determines whether an RG trajectory
connects to the fixed point in the UV or IR, in which case it is called relevant or irrelevant, respectively. 
While there are three irrelevant eigendirections, only one RG trajectory reaches the fixed point in the UV.
Thus, asymptotic safety is  established along a one-dimensional submanifold in parameter space. 
Emanating from the UV fixed point, the two outgoing trajectories lead either to IR freedom, or towards a strongly coupled regime with  either confinement or an interacting conformal fixed point.
For $\epsilon > \epsilon_\text{max}$, the UV fixed point disappears and the theory  is described by an effective field theory in the UV, and a free theory in the IR; see \fig{Schematic} for an illustration of these features.

\section{Computing Beta Functions}\label{sec:computation}

It is the central aim of this work to find and study the renormalisation group flow for the theory~\eq{lag} at the complete 3NLO order in the conformal expansion, which corresponds to the  \texttt{433} approximation. 
It requires four-loop $\overline{\text{MS}}$ $\beta$ functions for the gauge coupling $g$, 
as well as the three-loop ones for the Yukawa coupling $y$, and the scalar quartics $u$ and $v$.
Generic $\beta$ functions for the gauge and Yukawa couplings  have been obtained in 
Refs.~\cite{Bednyakov:2021qxa,Davies:2021mnc} using Weyl consistency conditions 
at \texttt{432}~\cite{Poole:2019kcm},\footnote{Note that our model~\eq{lag} is CP-even and cannot generate an additive $\beta$ function to its topological angle; thus, the caveat raised in Ref.~\cite{Valenti:2022uii} regarding the Weyl 
consistency condition does not apply.
} while the fully general quartic $\beta$ functions 
are  available at two-loop order \cite{Machacek:1984zw,Luo:2002ti}. These results 
are conveniently accessible via software packages such as 
\texttt{RGBeta}~\cite{Thomsen:2021ncy} and \texttt{FoRGEr}~\cite{Steudtner:FoRGEr}.
Moreover, quartic and Yukawa contributions to the three-loop $\beta$ functions for the single- and double-trace quartic couplings $u$ and $v$ 
have been determined in Refs.~\cite{Steudtner:2020tzo,Steudtner:2021fzs}.
Therefore, the only missing pieces for a complete \texttt{433} analysis 
are the three-loop contributions to $\beta_u$ and $\beta_v$ containing gauge interactions. Their computation is the main task of this section.

\subsection{Computational Strategy}
We have conducted a complete computation of all scalar, fermion, vector-boson, and 
ghost two-point functions, gauge and Yukawa vertex three-point functions, 
and scalar four-point functions up to three-loop order. 
This allows to compute the $\overline{\text{MS}}$ counterterms that 
determine all $\gamma$ and $\beta$ functions, including the missing three-loop 
results for the single- and double-trace quartic scalar couplings $\beta_{u,v}$.
While we are ultimately interested in the Veneziano limit, our computations 
have been conducted for finite $N_{f}$ and $N_c$.

The calculation has been achieved using the framework \texttt{MaRTIn}~\cite{MaRTIn},
which has been extended to three-loop order for this purpose.
All Feynman diagrams are generated using \texttt{QGRAF}~\cite{Nogueira:1991ex} 
and further evaluated in \texttt{FORM}~\cite{Kuipers:2012rf}.
Overall, almost 33,500 diagrams have been processed.
To distinguish UV and IR poles, we employ the technique of 
infrared rearrangement (IRA)~\cite{Misiak:1994zw,Chetyrkin:1997fm}.
For convenience, we choose the scalar mass in Eq.~\eq{lag} to be zero and 
expand each propagator (with integration momentum $p$) recursively with a universal mass parameter $m_\text{IRA}$
\begin{equation}\nonumber
    \frac{1}{(p - q)^2} = \frac{1}{p^2 - m_\text{IRA}^2} + \frac{2 \,p\!\cdot\! q - p^2 }{p^2 - m_\text{IRA}^2} \frac{1}{(p - q)^2} \,.
\end{equation} 
Finite terms with a sufficiently negative degree of divergence are dropped systematically. 
In order to cancel subdivergences in two- and three-loop diagrams, 
counterterms for scalar and vector-boson masses proportional to $m_{\text{IRA}}^2$ 
are introduced, while this is not necessary for ghosts or fermions~\cite{Chetyrkin:2012rz}.
In the end, all pole terms should be independent of $m_\text{IRA}$ and logarithms thereof, 
which is a non-trivial consistency check of the result.
We apply tensor and integration by parts reduction techniques \cite{Chetyrkin:1997fm} and the program \texttt{LiteRed}~\cite{Lee:2012cn,Lee:2013mka} is utilised to reduce all remaining three-loop scalar vacuum integrals to a set of masters \cite{Bobeth:1999mk,Martin:2016bgz}.

\subsection{Treatment of \texorpdfstring{$\gamma_5$}{gamma5}}
\begin{figure*}[ht]
    \centering
    \begin{tabular}{ccc}
         \begin{tikzpicture}
         \begin{feynhand}
         \vertex [particle] (u1) at (0em,+1em) {};
         \vertex [particle] (d1) at (0em,-1em) {};
         \vertex [dot] (u2) at (2em,1em) {};
         \vertex [dot] (d2) at (2em,-1em) {};
         \vertex [dot] (u3) at (4em,1em) {};
         \vertex [dot] (d3) at (4em,-1em) {};
         \vertex [dot] (u4) at (6em,1em) {};
         \vertex [dot] (d4) at (6em,-1em) {};
         \vertex [dot] (u5) at (8em,1em) {};
         \vertex [dot] (d5) at (8em,-1em) {};
         \vertex [particle] (u6) at (10em,1em) {};
         \vertex [particle] (d6) at (10em,-1em) {};
         \propag [scalar] (u1) to (u2);
         \propag [plain] (u2) to (u3);
         \propag [gluon] (u3) to (u4);
         \propag [plain] (u4) to (u5);
         \propag [scalar] (u5) to (u6);
         \propag [scalar] (d1) to (d2);
         \propag [plain] (d2) to (d3);
         \propag [gluon] (d3) to (d4);
         \propag [plain] (d4) to (d5);
         \propag [scalar] (d5) to (d6);
         \propag [plain] (u2) to (d2);
         \propag [plain] (u3) to (d3);
         \propag [plain] (u4) to (d4);
         \propag [plain] (u5) to (d5);
         \end{feynhand}
         \end{tikzpicture}
         &  
         \begin{tikzpicture}
         \begin{feynhand}
         \vertex [particle] (u1) at (0em,+1em) {};
         \vertex [particle] (d1) at (0em,-1em) {};
         \vertex [dot] (u2) at (2em,1em) {};
         \vertex [dot] (d2) at (2em,-1em) {};
         \vertex [dot] (u3) at (4em,1em) {};
         \vertex [dot] (d3) at (4em,-1em) {};
         \vertex [dot] (m4) at (6em,0em) {};
         \vertex [dot] (m6) at (8em,0em) {};
         \vertex [dot] (u5) at (7em,1em) {};
         \vertex [dot] (d5) at (7em,-1em) {};
         \vertex [particle] (u7) at (10em,1em) {};
         \vertex [particle] (d7) at (10em,-1em) {};
         \propag [scalar] (u1) to (u2);
         \propag [plain] (u2) to (u3);
         \propag [scalar] (d1) to (d2);
         \propag [plain] (d2) to (d3);
         \propag [plain] (u2) to (d2);
         \propag [plain] (u3) to (d3);
         \propag [scalar] (u5) to (u7);
         \propag [scalar] (d5) to (d7);
         \propag [plain] (m4) to [in=90, out=90, looseness=1.5] (m6);
         \propag [plain] (m4) to [in=-90, out=-90, looseness=1.5] (m6);
         \propag [gluon] (u3) to [out=0, in=165] (m6);
         \propag [gluon] (d3) to (m4);

         \end{feynhand}
         \end{tikzpicture}
         & 
         \begin{tikzpicture}
         \begin{feynhand}
         \vertex [particle] (u1) at (0em,+1em) {};
         \vertex [particle] (d1) at (0em,-1em) {};
         \vertex [dot] (m2) at (2em,0em) {};
         \vertex [dot] (u3) at (3em,1em) {};
         \vertex [dot] (d3) at (3em,-1em) {};
         \vertex [dot] (m4) at (4em,0em) {};
         \vertex [dot] (m5) at (6em,0em) {};
         \vertex [dot] (u6) at (7em,1em) {};
         \vertex [dot] (d6) at (7em,-1em) {};
         \vertex [dot] (m7) at (8em,0em) {};
         \vertex [particle] (u8) at (10em,1em) {};
         \vertex [particle] (d8) at (10em,-1em) {};
         \propag [scalar] (u1) to (u3);
         \propag [scalar] (d1) to (d3);
         \propag [scalar] (u6) to (u8);
         \propag [scalar] (d6) to (d8);
         \propag [plain] (m5) to [in=90, out=90, looseness=1.5] (m7);
         \propag [plain] (m5) to [in=-90, out=-90, looseness=1.5] (m7);
         \propag [plain] (m2) to [in=90, out=90, looseness=1.5] (m4);
         \propag [plain] (m2) to [in=-90, out=-90, looseness=1.5] (m4);
         \propag [gluon] (m4) to [out=15, in=165] (m5);
         \propag [gluon] (m2) to [out=-15, in=195] (m7);

         \end{feynhand}
         \end{tikzpicture}
    \end{tabular}
    \caption{Scalar four-point diagrams that fulfil Eq.~\eq{lc-requirement}, 
    but can still be treated without inconsistencies 
    within the na\"ive $\gamma_5$ scheme as argued in Ref.~\cite{Chetyrkin:2012rz} and discussed in the main text.
  \label{fig:quartic-gamma5-danger}}
\end{figure*}
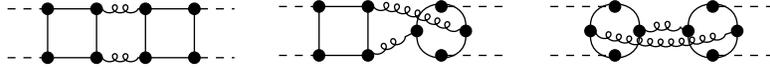

Moreover, we would like to comment on the treatment of $\gamma_5$, as its na\"ive definition
\begin{equation}\label{eq:naive-gamma5}
    \{\gamma_5, \gamma^\mu\} = 0, \qquad \gamma_5 = \frac{i}{4!} \varepsilon_{\mu \nu \rho \sigma}\, \gamma^\mu \gamma^\nu \gamma^\rho \gamma^\sigma
\end{equation}
with the 4-dimensional Levi-Civita symbol $\varepsilon$ is in conflict with the 
dimensional regularisation procedure.
In fact this treatment is algebraically inconsistent. 
In our case, the inconsistencies and ambiguities regarding the 
$\gamma_5$ treatment can only arise starting at three loops when contracting two different terms
$\propto  \mathrm{tr}\left(\gamma^\mu \gamma^\nu \gamma^\rho \gamma^\sigma \gamma_5\right)$ 
or with traces of more $\gamma$ matrices~\cite{Jegerlehner:2000dz}, e.g., from diagrams 
in \fig{quartic-gamma5-danger}.
As observed in Ref.~\cite{Poole:2019kcm}, such terms are only generated if 
for each closed fermion line $\ell$ 
with $n_g^{(\ell)}$ gauge-vertex insertions and $n_y^{(\ell)}$ Yukawa-vertex 
insertions 
\begin{equation}\label{eq:lc-requirement}
  2\,n_g^{(\ell)} + n_y^{(\ell)} \geq 5\,.
\end{equation}
This constraint cannot be satisfied for scalar two-point functions at three-loop order.
There is a single set of scalar four-point diagrams at three-loop where Eq.~\eq{lc-requirement} 
is fulfilled. These are diagrams containing two fermion loops with 
$n_y^{(1)} = n_g^{(1)} = n_y^{(2)} = n_g^{(2)} = 2$, as depicted in \fig{quartic-gamma5-danger}.
Each fermion loop in these diagrams features 
their own loop momentum, which can be integrated over independently from the rest of the diagram. External momenta can be set to zero as the $1/\varepsilon$ UV pole terms relevant for computing the $\beta$ functions are independent of them.
In the end, the $\gamma$ matrices along each fermion line either carry
Lorentz indices from the gauge boson propagators, or are  contracted with the third integration momentum 
exchanged between the loops.
In either case, there are insufficient independent Lorentz indices and momenta 
feeding into each fermion trace to form a tensor 
$\propto  \mathrm{tr}\left(\gamma^\mu \gamma^\nu \gamma^\rho \gamma^\sigma \gamma_5\right)$~\cite{Chetyrkin:2012rz}.
Hence, three-loop quartic $\beta$ functions, which represent the main novel result of our 
computation, cannot depend on the $\gamma_5$ scheme and can be treated without 
inconsistencies within the semi-na\"ive $\gamma_5$ scheme employed in this work and discussed below.
We extract the four-loop gauge and three-loop Yukawa $\beta$ functions
from literature results~\cite{Bednyakov:2021qxa,Davies:2021mnc}.
In this case it is known that all potential $\gamma_5$ ambiguities 
are fixed due to Weyl consistency conditions~\cite{Poole:2019txl,Poole:2019kcm,Davies:2019onf}.

In order to deal with $\gamma_5$ in our calculation, we employ the 
semi-na\"ive scheme~\cite{Chetyrkin:2012rz,Bednyakov:2012en} with
\begin{equation}
    \{\gamma_5, \gamma^\mu\} = 0, \qquad \gamma_5 = \frac{i}{4!} \widetilde{\varepsilon}_{\mu \nu \rho \sigma}\, \gamma^\mu \gamma^\nu \gamma^\rho \gamma^\sigma
\end{equation}
where $\widetilde{\varepsilon}$ is a $(4-2\varepsilon)$-dimensional, completely 
antisymmetric tensor which satisfies the identity 
\begin{equation}\label{eq:semi-naive}
    \widetilde{\varepsilon}^{\mu_1 \nu_1 \rho_1 \sigma_1} \widetilde{\varepsilon}_{\mu_2 \nu_2 \rho_2 \sigma_2} = - \delta^{[\mu_1}_{\;\;[\mu_2} \,\delta^{\nu_1}_{\;\; \nu_2} \,\delta^{\rho_1}_{\;\; \rho_2} \,\delta^{\sigma_1]}_{\;\; \sigma_2]} + \mathcal{O}(\varepsilon)\,.
\end{equation}
In exactly four space-time dimensions, $\widetilde{\varepsilon}$ is the Levi-Civita symbol 
and the na\"ive definition in Eq.~\eq{naive-gamma5} is recovered. 
Slightly away from the integer dimension at $d=4-2\varepsilon$, 
$\widetilde{\varepsilon}$ digresses by terms $\mathcal{O}(\varepsilon)$ from the Levi-Civita case. 
Hence, the otherwise four-dimensional identity in Eq.~\eq{semi-naive} picks up 
corrections $\mathcal{O}(\varepsilon)$.
The exact shape of these $\mathcal{O}(\varepsilon)$ corrections is irrelevant 
for the calculation of counterterms as long as Eq.~\eq{semi-naive} is only applied in 
terms that are already finite or only contain a single pole $\tfrac1{\varepsilon}$. 
We have verified that this is indeed the case in our calculation.
Finally, we would like to mention that poles due to non-hermitian field 
strength renormalisation tensors are absent as the flavour symmetry 
is unbroken~\cite{Bednyakov:2014pia,Herren:2017uxn, Jack:2016tpp,Herren:2021yur}.\\

\subsection{Consistency Checks}
Overall, the computation agrees at finite $N_{f,c}$ with generic literature results 
\cite{Machacek:1983tz,Machacek:1983fi,Machacek:1984zw,
Luo:2002ti,Pickering:2001aq,Mihaila:2014caa,Schienbein:2018fsw,Bednyakov:2021qxa,Davies:2021mnc} 
at \texttt{432} as well as previous calculations for \texttt{433} in the gaugeless 
limit~\cite{Steudtner:2020tzo,Steudtner:2021fzs}. 
To cross check the gauge contributions, we have extended the basis of tensor 
structures for the general scalar $\gamma$ and quartic $\beta$ 
functions~\cite{Steudtner:2021fzs} to account for gauge interactions 
among fermions (while retaining scalars as not charged). 
Details can be found in \App{template-rge}. Each tensor structure 
in the general $\beta$ functions has a universal coefficient that 
can be determined by comparing the corresponding renormalisation group equations of suitable 
literature results.
In our case, we have utilised the three-loop data for the Higgs 
self-interactions in the SM \cite{Chetyrkin:2012rz,Bednyakov:2013eba,Chetyrkin:2013wya,Bednyakov:2013cpa} 
with $g_1 = g_2 = 0$, as well as a QED-like gauge-Yukawa theory with a 
real scalar singlet \cite{Zerf:2018csr}. All references employ the same 
semi-na\"ive $\gamma_5$ scheme. 
The literature models are compatible with the generalised lagrangian \eq{master-template}. All relevant parts of their scalar quartic $\beta$ functions, mass and field anomalous dimensions can be computed using the prescription \eq{template-AD} and \eq{template-quart}, up to a number of model-independent coefficients. Comparing these results with the explicit computations of \cite{Chetyrkin:2012rz,Bednyakov:2013eba,Chetyrkin:2013wya,Bednyakov:2013cpa,Zerf:2018csr} yields relations of those coefficients.
Although not all coefficients can be fixed, the data is sufficient to obtain the complete  quartic $\beta$ functions for the theory \eqref{eq:lag}  by using the formalism of \eq{template-AD} and \eq{template-quart}.
We find full agreement with our explicit calculation at finite $N_{f,c}$. 

\subsection{Higher Orders}

In order to advance the conformal expansion  to 4NLO (\texttt{544} approximation), the complete 5-loop gauge as well as 4-loop Yukawa and quartic $\beta$ functions are required. Partial results are available from QCD-like theories~\cite{Gracey:1996he,Luthe:2016ima,Baikov:2016tgj,Herzog:2017ohr,Luthe:2017ttg,Chetyrkin:2017bjc}, and from purely scalar theories~\cite{Steudtner:2020tzo,Bednyakov:2021ojn}. What is missing, however, are the crucial contributions from  Yukawa interactions, the coupling that mediates between the gauge and scalar sectors. 
It is well-known that Yukawa interactions are key for the primary existence of the fixed point~\cite{Litim:2014uca,Bond:2016dvk,Bond:2018oco,Steudtner:2020jcj}, 
and their contributions are therefore expected to be equally important at higher orders.

As \texttt{544} requires the computation of 4-point functions, it is prudent to employ infrared rearrangement by massive propagators as demonstrated in this work.
Some tools for this endeavour have already been developed, see for instance~\cite{Laporta:2002pg,Schroder:2002re,Czakon:2004bu,Schroder:2005va,Luthe:2015ngq,Pikelner:2017tgv} and references therein.
However, and given the limitations of  the semi-na\"ive algorithm, the main new complication will be the consistent treatment of $\gamma_5$. Notice that up until now this has not been an issue in QCD-like or purely scalar theories.  
Also, while at \texttt{432} all $\gamma_5$-ambiguities have been removed using Weyl consistency conditions~\cite{Poole:2019txl,Poole:2019kcm}, it is far from  evident that the same can be achieved at higher orders. For starters, this would require the formulation of a basis for generalised \texttt{543} and \texttt{654}  $\beta$ functions, which in itself is a massive undertaking. We also point out that  a complete basis for the general quartic $\beta$ function at three loops does not yet exist. These ambitious endeavours are left for future work.

\section{Results\label{sec:betafunction}}

In this section, we summarise our results for $\beta$ functions and anomalous dimensions, and determine fixed points and universal scaling dimensions up to the third non-trivial order in the Veneziano parameter. We also discuss  aspects of unitarity, bounds on the conformal window, and the phase diagram and  UV-IR connecting trajectories in comparison with asymptotic freedom.
 \subsection{Beta Functions}
In this section we list the $\beta$ functions in the loop expansion
\begin{equation}
\beta_i \equiv \frac{\mathrm{d} \alpha_i}{\mathrm{d} \ln{\mu}} =  \sum_{\ell=1}^\infty \beta_i^{(\ell)},
\end{equation}
with $i= g,\,y,\,u,\,v$. The new pieces with respect to the previous analysis \cite{Bond:2017tbw} 
are the four loop contributions to the gauge   $\beta_g^{(4)}$, the three loop contribution to the Yukawa  $\beta_y^{(3)}$, and the three loop contributions to the scalar $\beta$ functions $\beta_{u,v}^{(3)}$. Specifically, 
\begin{widetext}

\begin{equation}\label{eq:beta-g}
  \begin{aligned}
    \beta_g^{(1)} &= \tfrac43 \epsilon\, \alpha_g^2 \,,\\[0.4em]
    \beta_g^{(2)} &= \left(25 + \tfrac{26}3 \epsilon\right) \alpha_g^3 - \tfrac12 \left(11 + 2 \epsilon\right)^2 \alpha_y \alpha_g^2\,,\\[0.4em]
    \beta_g^{(3)} &= \left(\tfrac{701}6 + \tfrac{53}3 \epsilon - \tfrac{112}{27} \epsilon^2 \right) \alpha_g^4 - \tfrac{27}8 \left(11 + 2 \epsilon\right)^2 \alpha_y \alpha_g^3 + \tfrac14\left(20 + 3 \epsilon\right) \left(11 + 2 \epsilon\right)^2\alpha_y^2 \alpha_g^2 \,,\\[0.4em]
    \beta_g^{(4)} &= -\left[\tfrac{14731}{72} + 550 \zeta_3 + \left(\tfrac{123473}{324} + \tfrac{1808}9 \zeta_3 \right) \epsilon + \left(\tfrac{21598}{243} + \tfrac{56}3 \zeta_3 \right)\epsilon^2 + \tfrac{260}{243} \epsilon^3\right] \alpha_g^5 \\
    &\phantom{ = } \ + \tfrac1{48} \left(11 + 2 \epsilon\right)^2  \left[\left(-107 + 432 \zeta_3  + \tfrac{758}3 \epsilon \right) \alpha_y \alpha_g^4 + 3\left(647 - 48 \zeta_3 + 92 \epsilon\right)\alpha_y^2 \alpha_g^3\right] \\
    &\phantom{ = } \  + \left(11 + 2 \epsilon\right)^2 \left[ 3 \,\alpha_u^2  - \left(\tfrac{875}{16} + \tfrac{179}{12} \epsilon + \tfrac{11}{12} \epsilon^2\right)\alpha_y^2 \right]\alpha_y\alpha_g^2 - \tfrac54 \left(11 + 2 \epsilon\right)^3 \alpha_u \alpha_y^2 \alpha_g^2\,.
  \end{aligned}
\end{equation}
We note that irrational coefficients  $\propto \zeta_3$ arise for the first time at four loop.
Further, the quartic coupling $\alpha_u$ makes its first appearance  at four loop, as it must.\footnote{Had the scalars been charged under the gauge symmetry, contributions would have appeared at three loop.}
This influence of the scalar sector is channelled through the Yukawa sector,
which itself is supplemented by three-loop results 
\begin{equation}\label{eq:beta-y}
  \begin{aligned}
   \beta_y^{(1)} &= \left(13 + 2 \epsilon\right) \alpha_y^2 - 6\,\alpha_g \alpha_y\,,\\[0.4em]
   \beta_y^{(2)} &= - \tfrac18 \left(35 + 2 \epsilon\right)\left(11 + 2 \epsilon\right) \alpha_y^3 + \left(49 + 8 \epsilon\right) \alpha_g \alpha_y^2   \\
   &\phantom{ = } \ - 4\left(11 + 2 \epsilon\right) \alpha_u \alpha_y^2 - \tfrac16 \left(93 - 20 \epsilon\right) \alpha_g^2 \alpha_y  + 4\,\alpha_u^2 \alpha_y\,,\\[0.4em]
   \beta_y^{(3)} &= \left(\tfrac{17413}{64} + \tfrac{2595}{32} \epsilon + \tfrac{59}{16} \epsilon^2  - \tfrac38 \epsilon^3\right) \alpha_y^4 - \tfrac12\left(118 + 19 \epsilon\right)\left(11 + 2 \epsilon\right) \alpha_g \alpha_y^3 \\
   &\phantom{ = } \ + 6\left(8 + \epsilon\right)\left(11 + 2 \epsilon\right) \alpha_u \alpha_y^3 - \left[\tfrac{1217}{16} + 198 \zeta_3  + \tfrac18 \epsilon \left(893 + 288 \zeta_3 + 136 \epsilon\right)\right] \alpha_g^2 \alpha_y^2 \\
   &\phantom{ = } \ + 2 \left(11 + 2 \epsilon\right) \alpha_g \alpha_u \alpha_y^2 + 5 \left(\tfrac52 +  \epsilon\right) \alpha_u^2 \alpha_y^2 - 8 \, \alpha_u^3 \alpha_y 
   \\ &\phantom{ = } \  
   + \left[\tfrac{641}6 + 132 \zeta_3 + \tfrac\epsilon {27} \left(1947 + 648 \zeta_3 + 70 \epsilon\right)\right] \alpha_g^3 \alpha_y\,.
  \end{aligned}
\end{equation}
In the quartic sector, the gauge dependent terms $\propto \alpha_g$ and 
$\propto \alpha_g^2$ 
are computed here for the first time.
This terms must vanish for $\alpha_y = 0$, which decouples the fermionic from the gauge sector.
This is indeed manifest in the evolution of both the single- and double-trace quartics.
The latter reads
\begin{equation}\label{eq:beta-u}
  \begin{aligned}
    \beta_u^{(1)} &= 8\,\alpha_u^2  + 4 \, \alpha_y \alpha_u - \left(11 + 2 \epsilon\right)  \alpha_y^2 \,,\\[0.4em]
    \beta_u^{(2)} &= - 24 \, \alpha_u^3 - 16 \,\alpha_y \alpha_u^2 - 3 \left(11 + 2 \epsilon\right) \alpha_y^2 \alpha_u 
    + 10\, \alpha_g \alpha_y \alpha_u  + \left(11 + 2 \epsilon\right)^2 \alpha_y^3 - 2 \left(11 + 2 \epsilon\right) \alpha_g \alpha_y^2 \\[.4em]
    \beta_u^{(3)} &=  104\, \alpha_u^4 + 34\, \alpha_u^3 \alpha_y  + \left(889 + 166 \epsilon\right) \alpha_u^2 \alpha_y^2  - \tfrac1{8}\left(\tfrac{11}2 + \epsilon\right)^2 (21 - 26 \epsilon) \alpha_y^4 \\
    &\phantom{ = } \ - \left(\tfrac{2953}{16} + \tfrac{315}8 \epsilon \right) \left(11 + 2\epsilon\right)\alpha_u \alpha_y^3 - (102 - 96 \zeta_3) \alpha_u^2 \alpha_y \alpha_g \\
     &\phantom{ = } \  + \tfrac14 \left(11 + 2 \epsilon\right) \left(149 - 240\zeta_3 \right) \alpha_u \alpha_y^2 \alpha_g 
     -\tfrac14 \left(11+ 2 \epsilon\right)^2 (5 - 24 \zeta_3) \alpha_y^3 \alpha_g \\
     &\phantom{ = } \ + \left(\tfrac{13}4 - 8 \epsilon\right) \alpha_u \alpha_y \alpha_g^2 + \tfrac18 \left(11 +2 \epsilon\right) \left(23 + 20 \epsilon\right) \alpha_y^2 \alpha_g^2\,.
  \end{aligned}
\end{equation}
Note the absence of a term $\propto \alpha_y \alpha_g^3$, which can easily be 
understood diagrammatically. 
As expected in the planar large-$N$ limit~\cite{Pomoni:2009joh}, the double-trace 
quartic does not enter other $\beta$ functions than its own, namely
\begin{equation}
  \begin{aligned}\label{eq:beta-v}
    \beta_v^{(1)} &= 12\,\alpha_u^2 + 16 \, \alpha_u \alpha_v + 4\, \alpha_v^2 + 4\, \alpha_y \alpha_v\,,\\[.4em]
    \beta_v^{(2)} &= - 96\,\alpha_u^3 - 40\, \alpha_u^2 \alpha_v - 24\, \alpha_y \alpha_u^2 - 32\, \alpha_y \alpha_u \alpha_v - 8\,\alpha_y \alpha_v^2 + 4\left(11 + 2 \epsilon\right) \alpha_u \alpha_y^2 \\ 
    &\phantom{ = } \  - 3 \left(11 + 2 \epsilon\right) \alpha_v \alpha_y^2 + 10\, \alpha_g \alpha_y \alpha_v + \left(11 + 2 \epsilon\right)^2 \alpha_y^3\,,\\[.4em]
     \beta_v^{(3)} &=  12\, \alpha_v^2 \alpha_u^2 + 480\, \alpha_v \alpha_u^3 + (772 + 384 \zeta_3 ) \alpha_u^4  + 66\,\alpha_v \alpha_u^2 \alpha_y + 192\,\alpha_u^3 \alpha_y \\ 
     &\phantom{ = } \ + \left(\tfrac{427}2 + 41 \epsilon\right) \alpha_v^2 \alpha_y^2 + \left(788 + 152 \epsilon + 96\zeta_3 \left(\tfrac{11}2 + \epsilon\right)\right) \alpha_v \alpha_u \alpha_y^2 \\
     &\phantom{ = } \ + \left(\tfrac{1985}2 + 187 \epsilon + 192\zeta_3 \left(\tfrac{11}2 + \epsilon\right)\right) \alpha_u^2 \alpha_y^2 \\
     &\phantom{ = } \ - 4 \left(\tfrac{11}2 + \epsilon\right) \left(105 + 22 \epsilon + 24 \zeta_3 \left(\tfrac{11}2 + \epsilon\right)\right) \alpha_u \alpha_y^3\\
     &\phantom{ = } \ - \tfrac1{8} \left(\tfrac{11}2 + \epsilon\right)\left(1545  + 374 \epsilon\right) \alpha_v \alpha_y^3  - \left(\tfrac{11}2 + \epsilon\right)^2 \left(73 + 10 \epsilon \right) \alpha_y^4 \\
     &\phantom{ = } \ - 9\left(17 - 16 \zeta_3\right) \alpha_u^2 \alpha_y \alpha_g - \left(204 - 192 \zeta_3\right) \alpha_v \alpha_u \alpha_y \alpha_g - \left(51 - 48 \zeta_3\right) \alpha_v^2 \alpha_y \alpha_g \\
     &\phantom{ = } \ + 8 \left(11 + 2 \epsilon\right) \left(7 - 9 \zeta_3\right) \alpha_u \alpha_y^2 \alpha_g + \tfrac14 \left(11 + 2 \epsilon\right) \left(149 - 240 \zeta_3\right) \alpha_v \alpha_y^2 \alpha_g \\
     &\phantom{ = } \ + \tfrac12 \left(11+2\epsilon\right)^2 \left(-1 + 12 \zeta_3\right) \alpha_y^3 \alpha_g + \left(\tfrac{13}4 - 8 \epsilon\right) \alpha_v \alpha_y \alpha_g^2 + 6 \left(11+2\epsilon\right)^2 \alpha_y^2 \alpha_g^2\,,
  \end{aligned}
\end{equation}
where it only appears to order $\propto \alpha_v^2$. 
This potentially leads to pair-wise fixed-point solutions 
that only differ by $\alpha_v^*$ and potentially merge at some value of $\epsilon$, 
disappearing into the complex plane.
Finite-$N$ corrections to \eq{beta-g}--\eq{beta-v} are more lengthy and can be found in \App{finite-N-betas}.

\end{widetext}

\subsection{Anomalous Dimensions}
Next, we provide some results for physical meaningful anomalous dimensions for mass and field strength renormalisation.
The scalar squared mass $m^2$ in~\eq{lag} corresponds to the only bilinear field operator  that does not violate local or global symmetries.
Its gauge-independent anomalous dimension
\begin{equation}
    \gamma_{m^2} = \frac{\mathrm{d} \ln m^2}{\mathrm{d}\ln \mu} = \sum_{\ell = 1}^\infty \gamma_{m^2}^{(\ell)} \,,
\end{equation}
cannot be obtained from our loop computation, since we have chosen $m^2=0$ for convenience. Thus, its counterterm is tainted 
by contributions from the gauge boson IRA mass.
Instead, we make use of the general $\beta$ functions for the quartic interactions, 
and employ the dummy field trick~\cite{Martin:1993zk,Luo:2002ti,Schienbein:2018fsw} 
to obtain mass $\beta$ functions. 
At three loops, we rely on the ansatz detailed in \App{template-rge} 
of tensor structures with the incomplete set of coefficients extracted from 
the literature, see \Sec{computation}. The information is sufficient to obtain
\begin{equation}\label{eq:gamma_mphi}
    \begin{aligned}
        \gamma_{m^2}^{(1)} &= 8\,\alpha_u + 4\,\alpha_v + 2\, \alpha_y\,,\\[0.4em]
        \gamma_{m^2}^{(2)} &= - 20\, \alpha_u^2 - 8\, \alpha_v \alpha_y - 16 \, \alpha_u \alpha_y + 5\,\alpha_y \alpha_g \\
        &\phantom{=}\  - \tfrac32(11+2\epsilon) \alpha_y^2\,,\\[0.4em]
        \gamma_{m^2}^{(3)} &= 240\,\alpha_u^3 + 12\,\alpha_v \alpha_u^2 + 33\, \alpha_u^2 \alpha_y \\
        &\phantom{=}\
        +\tfrac12(427 + 82 \epsilon) \alpha_v \alpha_y^2\\
        &\phantom{=}\
        + (394 + 264 \zeta_3 + 76 \epsilon + 48 \zeta_3 \epsilon) \alpha_u \alpha_y^2  \\
        &\phantom{=}\
          + 3 (16 \zeta_3 - 17) (2\,\alpha_u + \alpha_v) \alpha_y \alpha_g 
        \\
        &\phantom{=}\ - \tfrac1{32}(11 + 2 \epsilon) (1545 + 374 \epsilon) \alpha_y^3 \\
        &\phantom{=}\ 
        - \tfrac{1}{8} (11+2\epsilon)(240 \zeta_3 - 149) \alpha_y^2 \alpha_g  \\
        &\phantom{=}\ 
        + \tfrac18(13-32 \epsilon) \alpha_y \alpha_g^2 
         \,,
    \end{aligned}
\end{equation}
while anomalous dimensions of other scalar bilinear operators violating global symmetries cannot be determined.

As for the fermions, a Dirac mass term $ m_\psi\,\overline{\psi}\psi$ breaks the global symmetry but leaves the gauge symmetry intact. Its anomalous dimension can be extracted from the generic Yukawa $\beta$ function up to three-loops~\cite{Bednyakov:2021qxa,Davies:2021mnc}, again using  the dummy field trick. Employing the notation
\begin{equation}
    \gamma_{m_\psi} = \frac{\mathrm{d} \ln m_\psi}{\mathrm{d}\ln \mu} = \sum_{\ell = 1}^\infty \gamma_{m_\psi}^{(\ell)} \,,
\end{equation}
the results read
\begin{equation}\label{eq:gamma_mpsi}
    \begin{aligned}
        \gamma_{m_\psi}^{(1)} &= \tfrac{1}{2}(11+2 \epsilon) \alpha_y - 3\,\alpha_g\,,\\[0.4em]
        \gamma_{m_\psi}^{(2)} &= -\tfrac1{16} (11+2\epsilon)(23+2\epsilon) \alpha_y^2
        + 2 (11+2\epsilon) \alpha_y \alpha_g \\
        &\phantom{=}\ 
        - \tfrac{1}{12}(93-20\epsilon) \alpha_g^2\,,\\[0.4em]
        \gamma_{m_\psi}^{(3)} &= -\tfrac{11}{4} (11+2\epsilon) \alpha_u^2 \alpha_y + (11+2\epsilon)^2\alpha_u\alpha_y^2 \\
        &\phantom{=}\ 
        + \left(\tfrac{13387}{128} + \tfrac{2119}{64} \epsilon + \tfrac{49}{32} \epsilon^2 - \tfrac{3}{16} \epsilon^3\right) \alpha_y^3 \\
        &\phantom{=}\ -\tfrac{1}{8} (11+2\epsilon) (477 - 48\zeta_3 + 76 \epsilon) \alpha_y^2 \alpha_g  \\
        &\phantom{=}\ -\tfrac{1}{16} (11+2\epsilon) (113 + 288 \zeta_3 + 136 \epsilon) \alpha_y \alpha_g^2 \\
        &\phantom{=}\ + \left(\tfrac{641}{12} + 66\zeta_3 + \tfrac{649}{18} \epsilon + 12 \zeta_3 \epsilon + \tfrac{35}{27} \epsilon^2 \right) \alpha_g^3\,.
    \end{aligned}
\end{equation}
Furthermore, a renormalisation procedure of all field $X$ has been conducted via the substitution
\begin{equation}
    X_\text{bare} = \sqrt{Z_X} \,X\,.
\end{equation}
These field strength renormalisation factors $Z_X$ contain counterterms and imply anomalous dimensions
\begin{equation}\label{eq:AD-definition}
    \gamma_{X} =  \frac{\mathrm{d} \ln \sqrt{Z_{X}}}{\mathrm{d}\ln \mu} = \sum_{\ell = 1}^\infty \gamma_{X}^{(\ell)} \,.
 \end{equation}
Note that all factors $Z_X$ are just multiplicative numbers as the global 
symmetries remain intact. This excludes any ambiguities stemming from 
antihermitian parts of anomalous-dimension 
matrices~\cite{Bednyakov:2014pia,Herren:2017uxn,Jack:2016tpp,Herren:2021yur}.
However, field strength anomalous dimensions $\gamma_X$ are in general gauge dependent and thus unphysical. The scalar field anomalous dimension $\gamma_\phi$ represents a notable exception, as its fixed point value is part of the CFT data. Unsurprisingly, we find it to be gauge independent up to three loop order
\begin{equation}\label{eq:gamma_phi}
    \begin{aligned}
        \gamma_\phi^{(1)} &= \alpha_y\,,\\[0.4em]
        \gamma_\phi^{(2)} &=  2\,\alpha_u^2 + \tfrac52\,\alpha_y \alpha_g - \tfrac34 \left(11 + 2\epsilon\right) \alpha_y^2\,,\\[0.4em]
        \gamma_\phi^{(3)} &= - 4 \, \alpha_u^3 - \tfrac{15}2 \alpha_u^2 \alpha_y + \tfrac52(11 + 2 \epsilon) \alpha_u \alpha_y^2 \\
        & \phantom{=}\ + \tfrac{1}{64}(183+10\epsilon)(11 + 2 \epsilon) \alpha_y^3 \\
        &\phantom{=}\ - \tfrac1{16} (48 \zeta_3 - 5) (11 + 2 \epsilon) \alpha_y^2 \alpha_g \\
        &\phantom{=}\ + \tfrac1{16} (13 - 32 \epsilon) \alpha_y \alpha_g^2 \,.
    \end{aligned}
\end{equation}
As for the other fields, anomalous dimensions in $R_\xi$ gauge are collected in \App{gauge-dep-ads}.

\subsection{Fixed Point}
With $\beta$ functions available at the complete \texttt{433} order, we are now in a position to determine interacting fixed points accurately up to complete cubic order in the Veneziano parameter $\epsilon$. Complete sets of coefficients up to  quadratic order have previously been found in \cite{Bond:2017tbw} (see also \cite{Litim:2014uca}). 

Using the expansion \eq{FP-exp},
and solving $\beta_{i}(\alpha_{j}^*) = 0$ systematically  as a power series in  $\epsilon$, we find
for the gauge coupling coefficients
\begin{equation}\label{eq:agn}
  \begin{aligned}
  \alpha_g^{(1)} &= \tfrac{26}{57}\,,\\
  \alpha_g^{(2)} &= 23 \tfrac{75245 - 13068 \sqrt{23}}{370386}\,,\\
  \alpha_g^{(3)} &= \tfrac{353747709269}{2406768228} - \tfrac{663922754}{22284891} \sqrt{23} + \tfrac{386672}{185193}\zeta_3\,.\\
  \end{aligned}
\end{equation}
Note that $\zeta_3$ arises for the first time in the cubic coefficient. Similarly, for the  Yukawa coupling we obtain
\begin{equation}
  \begin{aligned}
  \alpha_y^{(1)} &= \tfrac{4}{19}\,,\\
  \alpha_y^{(2)} &= \tfrac{43549 }{20577} - \tfrac{2300 }{6859} \sqrt{23} \,,\\
  \alpha_y^{(3)} &= \tfrac{2893213181}{44569782} - \tfrac{96807908}{7428297} \sqrt{23} + \tfrac{4576}{6859}\zeta_3\,.\\
  \end{aligned}
\end{equation}
The single- and double-trace quartic scalar couplings give rise to the coefficients 
\begin{align}
  \alpha_u^{(1)} &= \tfrac{\sqrt{23}-1}{19}\,,\nonumber\\
  \alpha_u^{(2)} &= \tfrac{365825 \sqrt{23}-1476577}{631028}\,,\\
  \alpha_u^{(3)} &= - \tfrac{5173524931447 \sqrt{23} - 24197965967251}{282928976136} - \tfrac{416 (\sqrt{23}-12)}{6859}\zeta_3\,\nonumber
\end{align}
and
  \begin{align}
  \alpha_v^{(1)} &= \tfrac{\sqrt{20 + 6 \sqrt{23}}-2\sqrt{23}}{19}\,,\nonumber \\
  \alpha_v^{(2)} &= \tfrac{-643330 \sqrt{23} + 2506816}{631028} + \tfrac{452563 \sqrt{23} - 1542518}{315514 \sqrt{20 + 6 \sqrt{23}}}\,,\nonumber\\
  \alpha_v^{(3)} &= \tfrac{442552351896048-249223363466258 \sqrt{23}}{282928976136 (307 + 60 \sqrt{23})}   \\
   & + \tfrac{(122834160737083 - 26761631049822 \sqrt{23}) \sqrt{20 + 6 \sqrt{23}}}{282928976136 (307 + 60 \sqrt{23})}\nonumber \\
   & + \tfrac{659988864 \zeta_3 (942 - 338 \sqrt{23} + 39 \sqrt{2(-529426 + 583581 \sqrt{23})}) }{282928976136 (307 + 60 \sqrt{23})}\,,\nonumber
\end{align}
respectively. Numerically, the expansions read
\begin{equation}\label{eq:FP-expand}
  \begin{aligned}
    \alpha_g^* &=  \phantom{+ }0.456\,\epsilon + 0.781\,\epsilon^2 + 6.610\, \epsilon^3 + 24.137\, \epsilon^4 \\
    \alpha_y^* &=  \phantom{+ }0.211\,\epsilon + 0.508\,\epsilon^2 + 3.322\, \epsilon^3 + 15.212\, \epsilon^4 \\
    \alpha_u^* &=  \phantom{+ }0.200\,\epsilon + 0.440\,\epsilon^2 + 2.693\, \epsilon^3 + 12.119\, \epsilon^4 \\
    \alpha_v^* &=  - 0.137\,\epsilon - 0.632\,\epsilon^2  - 4.313\, \epsilon^3 - 24.147\, \epsilon^4\,, 
  \end{aligned}
\end{equation}
where we have neglected subleading corrections $\propto \epsilon^5$. 
All  coefficients up to and including $\propto \epsilon^3$ remain unchanged even if higher loops are included.
To indicate the trend  beyond  the strict \texttt{433} approximation, we also   show  the incomplete next-order  coefficients $\propto \epsilon^4$ that will receive as-of-yet unknown corrections at order  \texttt{544}.
At  the preceding loop order \texttt{322}, for example, the incomplete contributions $\propto \epsilon^3$ accounted for  $60-85 $\% of the complete cubic coefficients at order \texttt{433} \cite{Bond:2017tbw}.
We note from \eqref{eq:FP-expand} that  corrections for all couplings at any order arise with the same sign, and that the cubic coefficients are almost an order of magnitude larger than the quadratic ones. 

Finally, we note that since $\beta_v$ is quadratic in $\alpha_v$ to any loop order in perturbation theory \cite{Pomoni:2009joh}, there also exists  a second fixed point solution in the double-trace sector with $\alpha^*_{v-}\le \alpha^*_v$ and the coordinates for $\alpha^*_{g,y,u}$  unchanged \cite{Litim:2014uca,Litim:2015iea}. This second solution, however, is  unphysical in that it leads to an unstable vacuum \cite{Litim:2014uca,Litim:2015iea}.

\subsection{Scaling Exponents}
Universal critical exponents are obtained as the eigenvalues of the stability matrix
\begin{equation}\label{eq:stabmat}
    M_{ij} = \left. \frac{\partial \beta_i}{ \partial \alpha_{j}} \right|_{\alpha = \alpha^*}
\end{equation}
and can equally be expanded as a power series in the Veneziano parameter,
\begin{equation}
    \vartheta_i = \sum_{n=1}^\infty \epsilon^n \vartheta_i^{(n)}.
\end{equation}
The stability matrix factorises because the double-trace coupling does not  couple back into the single-trace couplings in the Veneziano limit. 
Quantitatively, we then find a single relevant and three irrelevant eigenvalues,
\begin{equation}
\vartheta_1<0<\vartheta_2<\vartheta_3 <\vartheta_4\,,
\end{equation}
and the UV critical surface due to canonically marginal interactions is one dimensional, with $\vartheta_3$ the isolated eigenvalue for the double-trace quartic. 

For $\vartheta_1$, the expansion starts out at quadratic order and is accurate up to including the fourth order,
\begin{equation}\label{eq:theta1}
  \begin{aligned}
    \vartheta_1^{(1)} &= 0\,,\\
    \vartheta_1^{(2)} &= -\tfrac{104}{171}\,,\\
    \vartheta_1^{(3)} &= \tfrac{2296}{3249}\,,\\
    \vartheta_1^{(4)} &= \tfrac{1405590649319}{15643993482} - \tfrac{15630102884}{869110749}\sqrt{23} + \tfrac{1546688}{555579} \zeta_3 \,.
  \end{aligned}
\end{equation}
The irrelevant directions start out at linear order and are accurate up to the cubic order in $\epsilon$. We find
\begin{equation} \label{eq:theta2} 
  \begin{aligned}
    \vartheta_2^{(1)} &= \tfrac{52}{19}\,,\\
    \vartheta_2^{(2)} &= \tfrac{136601719 - 22783308 \sqrt{23}}{4094823}\,,\\
    \vartheta_2^{(3)} &= -\tfrac{119064152144668585}{117078859819806} + \tfrac{93098590593718400}{448802295975923} \sqrt{23}\,,
  \end{aligned}
\end{equation}
as well as 
  \begin{align}
    \vartheta_3^{(1)} &= \tfrac{8}{19}\sqrt{20 + 6 \sqrt{23}}\,,\nonumber \\
 \label{eq:theta3}     \vartheta_3^{(2)} &= \tfrac{4(-1682358 + 410611 \sqrt{23})}{157757  \sqrt{20 + 6 \sqrt{23}}}\,, \\ \nonumber
     \vartheta_3^{(3)} &= 2\tfrac{96845792758245 \sqrt{23} + 8579855232  (19847 + 6564 \sqrt{23} )\zeta_3 }{35366122017 (307 + 60 \sqrt{23}) \sqrt{20 + 6 \sqrt{23}}}\ \ \ \ \\ \nonumber 
     &\quad -2\tfrac{616512472540856 }{35366122017 (307 + 60 \sqrt{23}) \sqrt{20 + 6 \sqrt{23}}}\,,
  \end{align}
and finally
  \begin{align}
    \vartheta_4^{(1)} &= \tfrac{16}{19}\sqrt{23}\,,\nonumber \\
 \label{eq:theta4}   \vartheta_4^{(2)} &= -\tfrac{44492672}{1364941} + \tfrac{272993948}{31393643}\sqrt{23} \,,\\ \nonumber
    \vartheta_4^{(3)} &= \tfrac{2 (-174067504271892880236 + 37418532792608300581 \sqrt{23})}{278706225801048183}\,.
  \end{align}
Numerically, the expansion coefficients read
\begin{equation}\label{eq:crit-exponents}
  \begin{aligned}
    \vartheta_1 &= -  0.608\,\epsilon^2 + 0.707\,\epsilon^3 + 6.947\, \epsilon^4 + 4.825\, \epsilon^5  \\
     \vartheta_2 &=  \ \ 2.737\,\epsilon + 6.676\, \epsilon^2 + 22.120\, \epsilon^3 + 102.55\, \epsilon^4 \\
     \vartheta_3 &= \ \  2.941\,\epsilon + 1.041\, \epsilon^2 + 5.137\, \epsilon^3 - 62.340\, \epsilon^4 \\
     \vartheta_4 &= \ \  4.039\,\epsilon + 9.107\, \epsilon^2 + 38.646\, \epsilon^3 + 87.016\, \epsilon^4 \,.
  \end{aligned}
\end{equation}
up to subleading corrections in $\epsilon$. 
We recall that all coefficients up to order $\epsilon^4$ for $\vartheta_1$ and up to order $\epsilon^3$ for $\vartheta_{2,3,4}$ remain unchanged even if higher loops are included, and that the  new coefficients from the order \texttt{433} are about ${\cal O}(4-9)$ times larger than those from the preceding order  \texttt{322}. Once more, to indicate the trend  beyond \texttt{433}, we also   show  the incomplete next-order coefficient.

\subsection{Bounds from  Series Expansions}\label{sec:power-series}

Next, we exploit the expansions of fixed point couplings and  exponents to estimate the size of the conformal window $\epsilon\le \epsilon_{\rm max}$, focusing on the coefficients which are unambiguously determined up to order \texttt{433}.

In order to satisfy vacuum stability, the quartic couplings must obey the conditions $0\le \alpha_u$  and $0\le \alpha_w\equiv \alpha_u + \alpha_v$  \cite{Paterson:1980fc,Litim:2015iea}. 
The former  is always satisfied, as can be seen from \eqref{eq:FP-expand}. Using the exact fixed point couplings for the latter, we find the series expansion
\begin{equation}\label{eq:stab-FP}
   \alpha_w^*= 0.063 \, \epsilon - 0.192   \, \epsilon^2 - 1.620\,\epsilon^3  +{\cal O}(\epsilon^4)\,.
\end{equation}
Corrections  arise with a sign opposite to the leading term. 
At order $\epsilon^2$, this implies $\epsilon\le \epsilon_\text{max}$, with  $\epsilon_\text{max}
\approx 0.327$ \cite{Bond:2017tbw,Bond:2021tgu}. At order $\epsilon^3$, the bound tightens by more than a factor of two, $\epsilon_\text{max}
\approx 0.147$. 
As an estimate for higher order corrections in $\epsilon$, we also employ a Pad\'e resummation  by writing of \eqref{eq:stab-FP} as $\alpha_w^*= \frac{A \epsilon +B   \epsilon^2}{1+C \epsilon}$ (and similarly for other couplings).\footnote{Notice that the loop  order \texttt{433} is the first perturbative order where resummation techniques can be applied.} 
Using the Pad\'e  approximant suggests that higher order effects  tighten the constraint even further,  $\epsilon_\text{max}\approx 0.087$.
Overall, we conclude that the series expansion \eq{stab-FP} indicates a loss of vacuum stability in the  range
 \begin{equation} \label{eq:vstab-fp}
	\begin{aligned}
		\epsilon_\text{max} \approx 0.087 - 0.146 \,.
	\end{aligned}
\end{equation}

We now turn to the expansion of scaling exponents, \eqref{eq:crit-exponents}. From the explicit expressions up to order \texttt{433}, we notice that the series expansion for  exponents $\vartheta_2$, $\vartheta_3$ and $\vartheta_4$ are monotonous, with same-sign corrections to the leading order, at every order. However  the relevant scaling exponent $\vartheta_1$  has all higher-order contributions with a sign opposite to the leading one. An overall change of sign is indicative for a collision of the UV fixed point with an IR fixed point. We estimate $\epsilon_{\rm max}$ from solving $\vartheta_1=0$ and reproduce the result 
$\epsilon_\text{max}\approx 0.860$  at order $\epsilon^3$  \cite{Bond:2017tbw,Bond:2021tgu}. The newly established  coefficient  at order $\epsilon^4$ now tightens  the constraint  by roughly a factor of three,  $\epsilon_\text{max}\approx 0.249$. 
Using a Pad\'e approximant as before, we find an even tighter estimate   $\epsilon_\text{max} \approx 0.091$. 
Overall, the series expansion indicates
that the conformal window terminates due to a fixed point merger 
in the  range
 \begin{equation} \label{eq:max}
	\begin{aligned}
		\epsilon_\text{max} \approx 0.091 - 0.249 \,.
	\end{aligned}
\end{equation}

Next, we ask whether the fixed point can disappear due to a merger  in the double-trace sector, $\alpha^*_{v-}\to  \alpha^*_v$ \cite{Benini:2019dfy}.
If so, it implies a double-zero of $\beta_v$ and  the corresponding scaling exponent must vanish, $\vartheta_3=0$. 
However, the first three universal expansion coefficients have all the same sign, \eqref{eq:crit-exponents}, giving no hints for a zero at \texttt{433}. Also, computing the difference between the double-trace quartic couplings
$\Delta \alpha_v\equiv \alpha^*_v-\alpha^*_{v-}$ we find
\begin{equation}\label{eq:mergerV}
    \begin{aligned}
       \Delta \alpha_{v}^*&=0.735 \,\epsilon+ 0.570\,\epsilon^2 + 
 0.326\,\epsilon^3 +  {\cal O}(\epsilon^4)\,.
    \end{aligned}
\end{equation}
 The first three  expansion coefficients in \eqref{eq:mergerV} have all the same sign,  offering no hints for a zero
 for any $\epsilon>0$. 
We conclude that a merger in the double-trace sector is  not supported by the \texttt{433} data.

 Finally, we provide a rough estimate for the range in $\epsilon$ with perturbative control.
 Based on naive dimensional analysis with couplings scaled in units of natural loop factors \cite{Weinberg:1978kz}, as done here, we take the view that this regime is characterised by $0<|\alpha^*|\lesssim 1$.\footnote{We stress that this criterion is not rigorous, and  must  be confirmed with higher loops or non-perturbatively.}
We note that the expansion coefficients  \eqref{eq:FP-expand} of the single- (double-) trace couplings receive only positive (negative) contributions, implying $\alpha^*_{g,y,u}>0$ and $\alpha^*_v<0$, and that the tightest bound $\epsilon<\epsilon_\text{strong} $ arises from the gauge coupling.
 We find $\epsilon_\text{strong} \approx 0.877$ at order $\epsilon^2$, and   $\epsilon_\text{strong} \approx 0.457$ at order $\epsilon^3$. To estimate higher order effects in $\epsilon$, we once more use a Pad\'e approximant for the gauge coupling fixed point and find the tighter bound  $\epsilon_\text{strong} \approx 0.117$, suggesting an onset of strong coupling in the range
 \begin{equation} \label{eq:strong}
	\begin{aligned}
		\epsilon_\text{strong} \approx 0.117 - 0.457 \,.
	\end{aligned}
\end{equation}
We notice that  regimes with vacuum instability or a  fixed point merger are reached  before the theory becomes strongly coupled. Also, in all cases \eqref{eq:vstab-fp}, \eqref{eq:max},
\eqref{eq:strong}, the tightest parameter bound arises from the Pad\'e resummations, giving bounds of the same size as  obtained from the  \texttt{322} $\beta$ functions \cite{Bond:2017tbw}.

\begin{figure}
    \centering
\includegraphics[width=.4\textwidth]{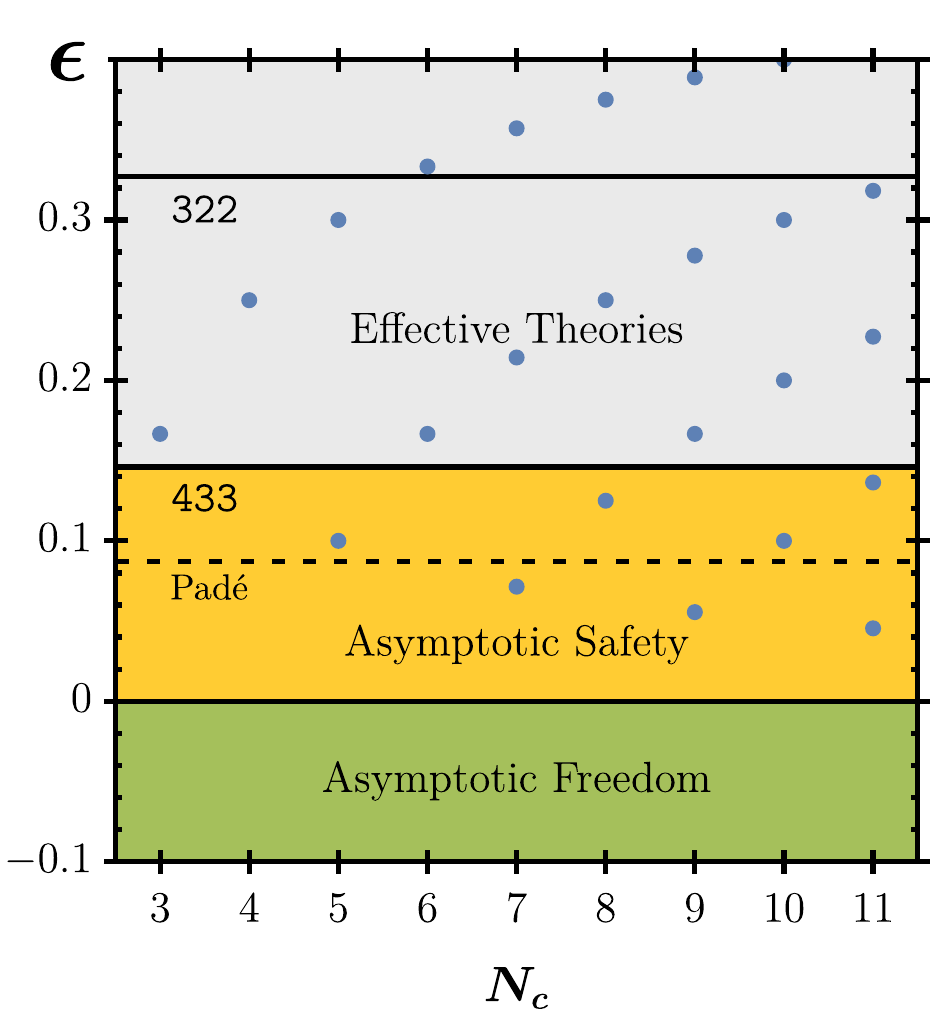}
    \caption{
The size of the UV conformal window  (yellow band) from series expansions, comparing  the new upper bound on $\epsilon$  at order  \texttt{433}, and the  Pad\'e approximant bound (dashed line), see \eqref{eq:vstab-fp}, with the previous upper bounds at order \texttt{322}   \cite{Bond:2017tbw}. Also shown are regimes with asymptotic freedom (green) and effective theories (grey). Dots indicate  integer values for $(N_c,N_f)$ in the $(\epsilon,N_c)$ plane.}
    \label{fig:433cw}
\end{figure}

In summary,  
the constraints on the conformal window as derived from series expansions of couplings  have become tighter,  owing to the  corrections established at order \texttt{433} over those at order \texttt{322}. The overall picture  shows that 
\begin{equation}\label{eq:ordering}
\epsilon_{\rm max}< \epsilon_{\rm strong}
\end{equation}
for each of the successive approximation orders \texttt{322}, \texttt{433}, and  for a Pad\'e approximant of the latter.
Results also  indicate that the conformal window is primarily limited  by the onset of vacuum instability and a nearby fixed point merger, rather than a merger in the double-trace sector or the onset of strong coupling.
Our results are further
illustrated in \fig{433cw}, including an extrapolation to finite field multiplicities $(N_c,N_f)$.
In particular, the smallest set of integer multiplicities compatible with an interacting UV fixed point increases from $(N_c,N_f)|_{\rm min}=(3,17)$  at  order \texttt{322} to   $(N_c,N_f)|_{\rm min}=(5,28)$ at  order \texttt{433}, and  to $(N_c,N_f)|_{\rm min}=(7,39)$  if we were to consider the forecast from Pad\'e approximants.
We defer a more detailed investigation of the conformal window 
to a forthcoming publication \cite{CW2023}.

\subsection{Unitarity}

In our setting, scale invariance at the weakly interacting UV fixed point  entails full conformal invariance \cite{Luty:2012ww}, and the critical theory can be described by a conformal field theory. Accordingly, the bound on unitarity for a spin-0 operator~$\mathcal{O}$ 
\begin{equation}\label{eq:unitarity}
    \Delta_\mathcal{O} = \text{dim}\ \mathcal{O} + \gamma_\mathcal{O}^* \geq 1
\end{equation}
must  be observed~\cite{Mack:1975je}. 
These CFT constraints can  be addressed by exploiting   our results for anomalous dimensions \eq{gamma_mphi} -- \eq{gamma_phi} and  fixed points \eq{FP-expand}. We find
\begin{equation}\label{eq:andimnum}
    \begin{aligned}
        \gamma_\phi^* &=   \phantom{-}0.2105 \, \epsilon  + 0.4625\, \epsilon^2 + 2.471\phantom{0}\, \epsilon^3 
        + \mathcal{O}(\epsilon^4)\,,\\
        \gamma_{m^2}^* &=   \phantom{-}1.470\phantom{0} \, \epsilon + 0.5207 \, \epsilon^2 + 2.568\phantom{0}\,\epsilon^3 
        + \mathcal{O}(\epsilon^4)\,,\\
        \gamma_{m_\psi}^* &=  - 0.2105\, \epsilon + 0.4628 \, \epsilon^2 + 0.3669 \, \epsilon^3 
        + \mathcal{O}(\epsilon^4)\,,
    \end{aligned}
\end{equation}
retaining all terms determined unambiguously in the $\epsilon$-expansion at \texttt{433}. Subleading terms starting at order $\epsilon^4$  necessitate the full \texttt{544} approximation.

We  observe from \eqref{eq:andimnum} that the scalar field and mass anomalous dimensions $\gamma_\phi^*$ and $\gamma_{m^2}^*$ are manifestly positive and  satisfy \eqref{eq:unitarity} without further ado. 
On the other hand, the fermion mass anomalous dimension $\gamma_{m_\psi}^*$ comes out negative to the leading order. Still, the subleading positive contributions up to cubic order  in $\epsilon$ ensure that the anomalous dimension remains strictly bounded from below, $\gamma_{m_\psi}^*\gtrsim -0.02$. In consequence, it cannot become sufficiently negative  for $ \Delta_\mathcal{\bar\psi\psi}$ to fall below the unitarity bound \eqref{eq:unitarity}. Altogether, we conclude that  the unitarity constraints \eqref{eq:unitarity} are satisfied non-marginally in perturbation theory. Moreover,  unitarity does not  offer bounds on $\epsilon$ within the conformal window.

\begin{figure*}
    \centering
    \includegraphics[width=.8\textwidth]{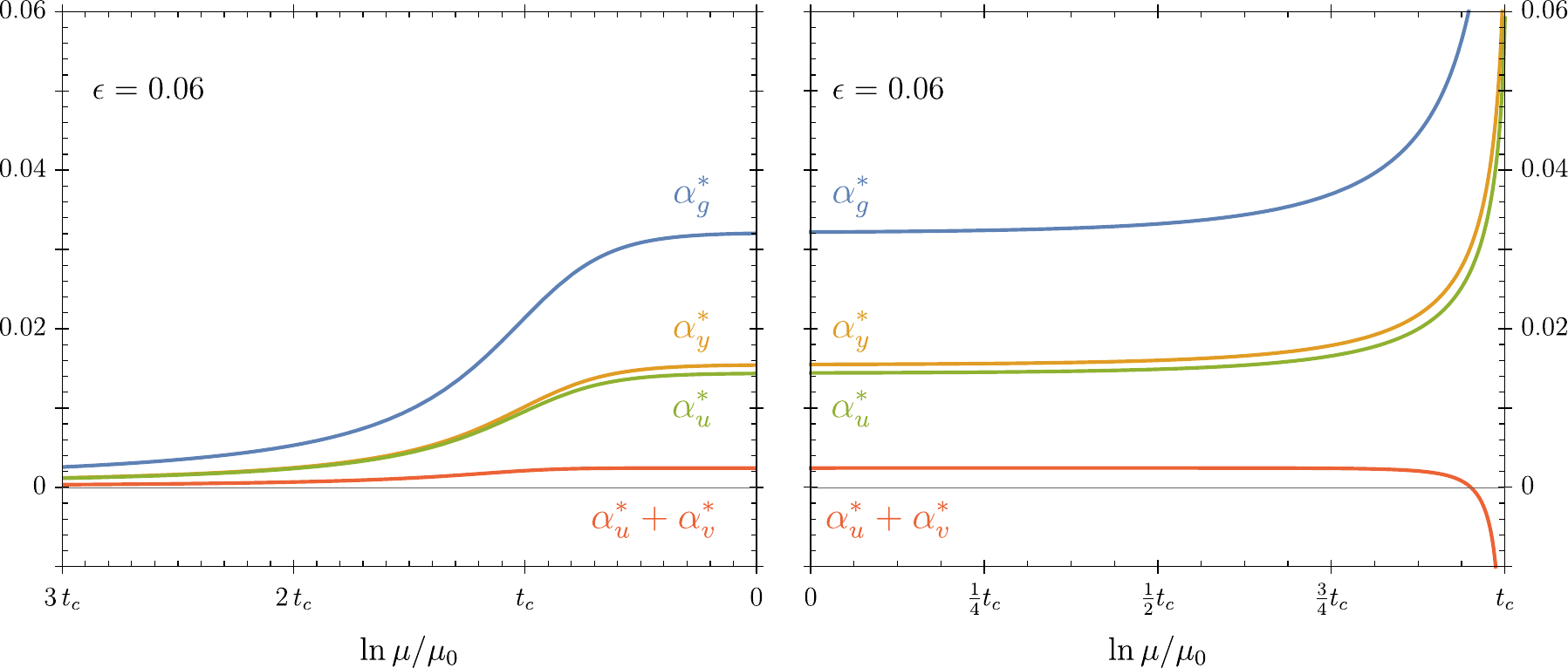}
    \caption{Running couplings along UV-IR connecting trajectories 
emanating out of the interacting UV fixed point, corresponding to  the (orange) separatrices highlighted in \fig{PD}.
Trajectories with UV initial condition $\delta\alpha_g(\mu_0)<0$ 
approach a free theory in the IR (left panel),
 while those with $\delta\alpha_g(\mu_0)>0$ enter a strongly coupled regime with  either confinement or conformality in the IR (right panel).
      Here, $t_c=\ln \Lambda_c/\mu_0$ with $\Lambda_c$ as in \eqref{eq:crossover1} and \eqref{eq:Lambdac}, respectively, and  $\epsilon = 0.06$.   }
    \label{fig:run-stream-combined}
\end{figure*}

\subsection{Scales and Phase Diagram}
We are now in a position to revisit the phase diagram of the theory. \fig{PD} illustrates the phase diagram  in the \texttt{433} approximation. Trajectories are shown in the $(\alpha_g,\alpha_y)$ plane, with arrows pointing from the UV to the IR. Evidently, asymptotic freedom is absent and the Gaussian fixed point is an IR attractive fixed point for all couplings. Nevertheless, the theory is UV-complete and remains predictive up to highest energies, courtesy of the interacting UV fixed point. It displays a single relevant direction amongst the classically marginal interactions. Without loss of generality, we take
\begin{equation}\label{eq:deltaa}
\delta\alpha_g=\alpha_g-\alpha_g^*
\end{equation}
as the fundamentally free parameter at the high scale $\mu_0$. The running of the Yukawa and quartic couplings $\alpha_i(\mu)$ with $i=y,u,v$ is entirely dictated by the running of $\alpha_g(\mu)$, and they can be expressed in terms of the gauge coupling as  $\alpha_i(\mu)=F_i[\alpha_g(\mu)]$ for suitable functions $F_i$.  The IR fate of trajectories emanating from the fixed point is determined by whether $\delta\alpha_g<0$ or $\delta\alpha_g>0$ at the high scale. In the former case, the theory becomes free in the infrared. In the latter case, the theory becomes strongly coupled and displays either confinement (such as in QCD), or IR conformality such as at an interacting IR fixed point. 
Our results are illustrated  in  \fig{run-stream-combined}, where sample trajectories connecting the UV fixed point with the IR are shown, also contrasting settings for initial conditions $\delta\alpha_g(\mu_0)<0$ (left panel) leading to IR freedom, with  initial conditions $\delta\alpha_g(\mu_0)>0$ (right panel).

The transition from the UV to the IR is characterised by an RG-invariant  scale $\Lambda_c$, analogous to $\Lambda_{\rm QCD}$ in QCD. It arises through dimensional transmutation  from  the dimensionless fundamental parameter $\delta\alpha_g\ll |\alpha_g^*|$ at the high scale $\mu$, and reads
\begin{equation}\label{eq:crossover1}
\Lambda_c \propto \mu\cdot \big|\delta\alpha(\mu)\big|^\nu\,,
\end{equation}
where $\nu = -1/\vartheta_1$ with $\vartheta_1$ the  relevant scaling exponent \eqref{eq:crit-exponents}. One readily confirms that ${\rm d}\Lambda_c/{\rm d}\ln\mu = 0$. The proportionality constant $c$ can be determined from a cross-over condition.  For $\delta\alpha_g>0$, strong coupling sets in as soon as $\delta\alpha_g$ is of order unity, hence $c\approx 1$.  For negative $\delta\alpha_g$, the Gaussian fixed point takes over as soon as $\delta\alpha_g\approx -\alpha_g^*/3$ \cite{Litim:2015iea},  giving $c=(3/\alpha_*)^\nu$ instead. 

We briefly  compare this with asymptotically free theories by taking $\epsilon<0$. In this case, the UV critical surface is now two-dimensional with both the gauge and the Yukawa coupling being marginally relevant. We then find a range of asymptotically free trajectories emanating from the Gaussian fixed point and characterised by
\begin{equation}\label{eq:deltaay}
\begin{aligned}
\delta\alpha_g&=\alpha_g-\alpha_g^*\\
\delta\alpha_y&=\alpha_y-\alpha_y^*\,.
\end{aligned}
\end{equation} 
For sufficiently small $0< -\epsilon\ll1$ and $\alpha_y=0$, the theory also displays a Banks-Zaks fixed point $\alpha_g^*$ of order $|\epsilon|$ and $\alpha_i^*=0 \ \ (i=y,u,v)$.
In either case, we find the transition   scale $\Lambda_c$ as 
\begin{equation}\label{eq:Lambdac}
\Lambda_c \propto \mu\cdot \exp\left(\frac1{\beta^{(1)}_g\,\delta \alpha(\mu)}\right)\,,
\end{equation}
characteristic for asymptotic freedom, with a negative one-loop gauge coefficient $\beta^{(1)}_g<0$. All trajectories  run towards  strong coupling where either confinement or conformality take over, except for the trajectory which terminates at the Banks-Zaks fixed point.  

Even though the UV critical surface is two-dimensional, it is interesting to note that the Yukawa nullcline is an IR attractor and  all outgoing trajectories collapse onto it. Hence, as soon as the  gauge coupling becomes of the order of the one-loop gauge coefficient, $\alpha(\mu)\gtrsim |\epsilon|$,  outgoing trajectories  along the nullcline of the asymptotically free theory (with $0<-\epsilon\ll 1$) become indistinguishable from  the outgoing trajectory  of the asymptotically safe theory (with $0<\epsilon\ll 1$).

\section{Discussion and Outlook}\label{sec:conclusion}

The quantum field theory \eqref{eq:lag} provides an important  template for an asymptotically safe 4d particle theory with an interacting and perturbatively-controlled fixed point at highest energies. 
We have extended the investigation of the UV theory up  to four loops in perturbation theory, a prerequisite to achieve the  complete cubic order in the underlying conformal  expansion in terms of a small Veneziano parameter $\epsilon$.
The central input for this are the  four-loop gauge,  three-loop Yukawa and quartic $\beta$ functions, and three-loop  anomalous dimensions. We have computed the  previously missing pieces which are the three-loop contributions to scalar $\beta$ functions containing gauge interactions 
  (\Sec{computation}).

With these results at hand, we have  determined all fixed point couplings, critical exponents, and anomalous dimensions up to the third non-trivial order in $\epsilon$, also investigating  the phase diagram and UV-IR connecting trajectories  (\Sec{betafunction}). 
Findings are  in accord with  unitarity, as they must. 
Most notably, bounds on the conformal window \eqref{eq:vstab-fp}, \eqref{eq:max}  have  become tighter  in comparison with the preceding order, and strengthen the view that the upper boundary remains under perturbative control.  
Our work further substantiates the existence of the fixed point at finite values of the Veneziano parameter and at finite $N$. 
Ultimately, conformality is lost 
due to the onset of vacuum instability and a nearby fixed point merger, 
\eqref{eq:ordering}, rather than through a  merger in the double-trace sector or strong coupling phenomena.

While our results have been achieved specifically for  Dirac fermions coupled to  $SU$ gauge fields and complex scalars $\phi_{ij}$ (see \tab{matter}), they equally hold true for 
theories with Majorana fermions coupled to  either  $SO$ gauge fields  with  symmetric complex scalars $\phi_{(ij)}$, or to  $Sp$ gauge fields with antisymmetric scalars $\phi_{[ij]}$
\cite{Bond:2019npq}. 
The reason for this is that these three types of matter-gauge theories are mutually equivalent to each other  in the Veneziano limit, even though   field content, gauge symmetries, and global symmetries are different \cite{Bond:2019npq}.
In particular, and modulo the normalisation of couplings,  their $\beta$ functions are identical to any loop order,
and  results  of this work  are equally valid for the partner theories.

We close with a few comments from the viewpoints of  lattice  Monte Carlo simulations, conformal field theory, and model building. 
It would be valuable  to  explore  the UV conformal window using complementary tools such as the lattice,   taking advantage of the vast body of works on  IR fixed points in 4d matter-gauge theories  \cite{Aarts:2023vsf}.
In a related vein, in our QFT setting, scale invariance at the UV fixed point  entails full conformal invariance \cite{Luty:2012ww}. Hence, our renormalisation group results offer  direct  access to conformal data characterising an interacting 4d CFT \cite{Cardy:1996xt,Codello:2017hhh}. 
It would then be equally important to investigate the 4d UV critical theory   using first-principle CFT  methods such as the bootstrap \cite{Li:2020bnb}, or other. Finally, we emphasise  that our setting provides a  blueprint for concrete 4d non-supersymmetric CFTs with standard model-like  field content in the UV, which  invites further  model building.
\\[4ex]

\centerline{\bf Acknowledgments}


We are thankful to Anders Eller Thomsen for  comments about parameterising $\gamma_5$ diagrams. Some of our results have been presented by one of us (TS) at {\it Asymptotic Safety meets Particle Physics IX}, Dec 2022, DESY, Hamburg, and at {\it Loopfest XXI}, June 2023, SLAC National Accelerator Laboratory.
This work is supported by the Science and Technology Facilities Council (STFC) under the Consolidated Grant ST/T00102X/1 (DL).

\appendix
\section{Tensor Structures for Three-Loop Quartic RGEs}\label{sec:template-rge}
In this appendix, we detail the general structure for three-loop scalar field anomalous dimensions and quartic $\beta$ functions of any renormalisable QFT with uncharged scalars. We closely follow the notation of \cite{Steudtner:2021fzs} with the Lagrangian
  \begin{equation}\label{eq:master-template}
  \begin{aligned}
    \mathcal{L} =
     & \  \tfrac{1}{2} \partial^\mu \phi_a \partial_\mu \phi_a + \tfrac{i}{2} \psi^j \tilde{\sigma}^\mu D_\mu \psi_j   -\tfrac{1}{4} \,g^{-2}_{AB}\,F_{\mu\nu}^A F^{B \mu\nu} 
    \\ & - \tfrac{1}{2} y^{ajk} \,\phi_a(\psi_j \varepsilon \psi_k)   - \tfrac{1}{24} \lambda_{abcd} \,\phi_a \phi_b \phi_c \phi_d\,\\
     &  - \tfrac{1}{2} \mathrm{m}^{jk} \,(\psi_j \varepsilon \psi_k)  
-\tfrac{1}{2} m^2_{ab} \, \phi_a \phi_b - \tfrac{1}{6} h_{abc} \,\phi_a \phi_b \phi_c\\
&+ \mathcal{L}_\text{gauge-fix}  + \mathcal{L}_\text{ghost} \,,
  \end{aligned}
  \end{equation}
 featuring real scalar field components $\phi^a$, vectors of fermionic Weyl components as well as their conjugates $\psi_i = (\psi^i)^*$ whose spinor indices are contracted by the two-dimensional Levi-Civita $\varepsilon$ as well as gauge fields with the field strength tensors $F^A_{\mu\nu}$ and covariant derivative $D_\mu$. Note the latter do not couple to scalars in accord with our model. Moreover, only the gauge coupling square $g^2_{AB}$, Yukawa interaction $y_{aij}$ and scalar quartic coupling $\lambda_{abcd}$ are relevant, while we will neglect fermionic $\mathrm{m}^{ij}$ and scalar masses $m^2_{ab}$ and cubic terms $h_{abc}$. In the same manner, details of the gauge fixing and Fadeev-Popov ghosts do not play a role. 
We will make fermionic indices implicit wherever appropriate. 
Moreover, $t^A_{ij}$ is introduced as the fermionic generator, and $(C_2^G)^{AB}$ as the Casimir invariant of the gauge interaction. Moreover, the fermion Casimir and Dynkin index are given by
\begin{equation}
      (C_2^F)_{ij} =  g^2_{AB} \, (t^A t^B)_{ij}, \qquad (S_2^F)^{AB} = \mathrm{tr}\left(t^A t^B\right) .
\end{equation}

The quartic $\beta$ function is composed of external leg and vertex corrections
\begin{equation}\label{eq:beta-lambda}
    \beta_{\lambda,3\ell}^{abcd} = \gamma_{\phi,3\ell}^{e(a} \,  \lambda^{bcd)e} + \beta_{\phi^4, 3\ell}^{(abcd)}
\end{equation}
The gaugeless part of both quantities are given in \cite{Steudtner:2021fzs}. In this limit, the leg corrections coincide with the scalar anomalous dimension. For gauge dependent terms, the bases agree only structurally. The reason is that each coefficient may also contain vertex corrections that can be brought to the shape of leg corrections using gauge transformations of tensor structures. For instance, this cancels all gauge dependence of the anomalous dimensions.
The gauge interaction terms missing in \cite{Steudtner:2021fzs} are of order $\propto y^4 g^2$ and $\propto y^2 g^4$ and read
\begin{widetext}
\begin{equation}\label{eq:template-AD}
 \begin{aligned}
    \gamma_{\phi,3\ell}^{ab} = \  & \gamma_{\phi,3\ell}^{ab} \big|_{g^2 = 0} \\   & 
    + G_{1} \, \mathrm{tr} \left(y^a C_2^F y^c y^b y^c + y^b C_2^F y^c y^a y^c \right) 
    + G_2\,\mathrm{tr} \left(y^a y^c C_2^F y^c y^b\right) 
    + G_3 \, \mathrm{tr}\left(y^a y^b y^c y^c C_2^F + y^b y^a y^c y^c C_2^F\right)  \\
    & +  G_4\, \mathrm{tr}\left(y^a C_2^F y^b y^c y^c\right) 
     + G_5\,\mathrm{tr}\left(y^a t^A y^b y^c t^B y^c\right) g^2_{AB} 
     + G_6 \, \mathrm{tr}\left(y^a C_2^F y^b C_2^F\right) \\ & 
     + G_7 \, \mathrm{tr}\left(y^a C_2^F C_2^F  y^b\right) 
     + G_8 \, \mathrm{tr}\left(y^a y^b t^A t^B\right) \left(g^2 S_2^F g^2\right)_{AB} 
     + G_9 \, \mathrm{tr}\left(y^a y^b t^A t^B\right) \left(g^2 C_2^G g^2\right)_{AB} \,.
 \end{aligned}
\end{equation}
Here $G_{1..9}$ are \textit{a priori} unknown coefficients that can be fixed by an actual loop computation. Note that this parametrisation assumes $\gamma_{\phi,3\ell}$ to be symmetric, as there is no explicitly broken flavour symmetry in our model.

As for the vertex corrections, we again refer to \cite{Steudtner:2021fzs} for the gaugeless tensor structures, and provide the missing ones:
\begin{equation}\label{eq:template-quart}
    \begin{aligned}
       \beta_{\phi^4, 3\ell}^{abcd} &=   \beta_{\phi^4, 3\ell}^{abcd}\big|_{g^2 = 0} + Q_1\,\lambda^{abef} \lambda^{cdeg}\, \mathrm{tr}\left(y^f y^g C_2^F\right)
       + Q_2\,\lambda^{abef}\,\mathrm{tr}\left(y^c y^e y^f y^d C_2^F\right) \\
       & 
       + Q_3\,\lambda^{abef}\,\mathrm{tr}\left(y^c y^e t^A y^d y^f t^B\right) g^2_{AB}
       + Q_4\,\lambda^{abef}\,\mathrm{tr}\left(y^c y^e t^A y^f y^d  t^B\right) g^2_{AB} \\
       & 
       + Q_5\, \lambda^{abef}\,\mathrm{tr}\left(y^c t^A y^e y^f t^B y^d  \right) g^2_{AB}
       + Q_6\, \lambda^{abef}\,\mathrm{tr}\left(y^c y^e C_2^F y^f y^d  \right) g^2_{AB} \\
       & 
       + Q_7\, \lambda^{abef}\,\mathrm{tr}\left(y^c y^e C_2^F y^d y^f  \right) g^2_{AB} 
       + Q_8\, \lambda^{abef}\,\mathrm{tr}\left(y^c y^e y^f C_2^F y^d  \right) g^2_{AB} \\
       & 
       + Q_{9}\, \mathrm{tr}\left(y^a y^b y^c y^d C_2^F C_2^F\right) 
       + Q_{10}\,\mathrm{tr}\left(y^a y^b y^c C_2^F y^d C_2^F\right)\\
       & 
       + Q_{11}\,\mathrm{tr}\left(y^a y^b C_2^F y^c y^d C_2^F\right)
       + Q_{12}\,\mathrm{tr}\left(y^a y^b t^A y^c  C_2^F y^d t^B\right) g_{AB}^2 \\
       &
       + Q_{13}\,\mathrm{tr}\left(y^a t^A C_2^F y^b y^c t^B y^d\right) g_{AB}^2
       + Q_{14}\,\mathrm{tr}\left(y^a t^A y^c t^C  y^b t^B y^d \right) g_{AB}^2 \, g_{CD}^2 \\
       &
       + Q_{15}\,\mathrm{tr}\left(y^a t^A t^C y^b t^B y^c t^D y^d\right) g_{AB}^2 \, g_{CD}^2\\
       &
        + Q_{16}\,\mathrm{tr}\left(y^a y^b t^A t^C y^c y^d t^B t^D\right) g_{AB}^2 \, g_{CD}^2 \\
       &
        + Q_{17}\, \mathrm{tr}\left(y^a y^b t^A t^C\right) \, \mathrm{tr}\left(y^c y^d t^B t^D\right)  g_{AB}^2 \, g_{CD}^2 \\
       & 
       + Q_{18}\,\mathrm{tr}\left(y^a t^A y^b t^C\right)\,\mathrm{tr}\left( y^c t^B y^d t^D \right) g_{AB}^2 \, g_{CD}^2 \\
       &
       + Q_{19}\,\mathrm{tr}\left(y^a t^A y^b t^C\right)\,\mathrm{tr}\left( y^c y^d t^B  t^D \right) g_{AB}^2 \, g_{CD}^2\\
       & 
       + Q_{20}\, \mathrm{tr}\left(y^a y^b t^A y^c y^d t^B\right) \left(g^2 S_2^F g^2\right)_{AB} 
       + Q_{21}\, \mathrm{tr}\left(y^a y^b t^A y^c y^d t^B\right) \left(g^2 C_2^G g^2\right)_{AB} \\
       & 
       + Q_{22}\, \mathrm{tr}\left(y^a y^b  y^c t^A y^d t^B\right) \left(g^2 S_2^F g^2\right)_{AB} 
       + Q_{23}\, \mathrm{tr}\left(y^a y^b y^c t^A y^d t^B\right) \left(g^2 C_2^G g^2\right)_{AB} \\
       & 
       + Q_{24}\, \mathrm{tr}\left(y^a y^b  y^c  y^d t^A t^B\right) \left(g^2 S_2^F g^2\right)_{AB} 
       + Q_{25}\, \mathrm{tr}\left(y^a y^b y^c  y^d t^A t^B\right) \left(g^2 C_2^G g^2\right)_{AB} \\
       &
       + Q_{26}\, \mathrm{tr}\left(y^a t^A y^b y^e y^c t^B y^d y^e \right)\,g_{AB}^2 
       + Q_{27}\, \mathrm{tr}\left(y^a y^b C_2^F y^c y^e y^d y^e \right) \\
       & 
       + Q_{28}\, \mathrm{tr}\left(y^a y^b y^e t^A y^c y^e y^d t^B \right)\,g_{AB}^2
       + Q_{29}\, \mathrm{tr}\left(y^a y^b t^A y^e y^c t^B y^d y^e \right)\,g_{AB}^2 \\
       &
       + Q_{30}\, \mathrm{tr}\left(y^a y^b y^e y^e y^c y^d C_2^F \right)
       + Q_{31}\, \mathrm{tr}\left(y^a y^b y^c y^e y^e y^d C_2^F \right) \\
       &
       + Q_{32}\, \mathrm{tr}\left(y^a t^A y^e y^b t^B y^c y^d y^e \right)\,g_{AB}^2
       + Q_{33}\, \mathrm{tr}\left(y^a y^b y^c y^e y^d y^e C_2^F \right) \\
       &
       + Q_{34}\, \mathrm{tr}\left(y^a y^b y^e y^c y^d y^e C_2^F \right)
       + Q_{35}\, \mathrm{tr}\left(y^a y^b y^e y^c C_2^F y^d y^e \right) \\
       &
       + Q_{36}\, \mathrm{tr}\left(y^a y^b t^A y^c y^e y^e y^d t^B \right)\,g_{AB}^2
       + Q_{37}\, \mathrm{tr}\left(y^a t^A y^e y^b y^c t^B y^e y^d\right)\,g_{AB}^2 \\
       &
       + Q_{38}\, \mathrm{tr}\left(y^a t^A y^b y^c y^e t^B y^e y^d\right)\,g_{AB}^2
       + Q_{39}\, \mathrm{tr}\left(y^a t^A y^e y^e y^b y^c t^B y^d\right)\,g_{AB}^2 \\
       &
       + Q_{40}\, \mathrm{tr}\left(y^a y^b y^c y^e C_2^F y^d y^e\right)
       + Q_{41}\, \mathrm{tr}\left(y^a y^b y^c t^A y^d y^e t^B y^e\right)\,g_{AB}^2 \\
        &
       + Q_{42}\, \mathrm{tr}\left(y^a y^b y^c y^d C_2^F y^e y^e \right)
       + Q_{43}\, \mathrm{tr}\left(y^a y^b y^c y^d y^e C_2^F y^e \right)
    \end{aligned}
\end{equation}
where $Q_{1..43}$ are again open coefficients. Note that we do not need to account for any non-na\"ive influence of $\gamma_5$ at this loop order as discussed in \Sec{computation}.

\section{Finite-\texorpdfstring{$N$}{N} Beta Functions}\label{sec:finite-N-betas}

Here we present the finite-$N$ corrections to the $\beta$ functions \eq{beta-g}--\eq{beta-v}. Apart from the Veneziano parameter $\epsilon$, these also retain an explicit dependence on inverse powers of the parameter $N_c$.
An extensive analysis of the finite-$N$ conformal window at 2NLO was conducted in \cite{Bond:2021tgu}, see there for explicit expressions up to \texttt{322}.
In the following, we make use of the abbreviations $r_c \equiv N_c^{-2}$ and $r_f \equiv \left[(\tfrac{11}2 + \epsilon) N_c\right]^{-2}$ and provide the four loop gauge $\beta$ function
\begin{equation}
  \begin{aligned}
    \alpha_g^{-2}\, \beta_g^{(4)} &= \big\{ \left[ -\tfrac{260 }{243}\epsilon^3 + \left(-\tfrac{56 \zeta_3}{3}-\tfrac{21598}{243}\right) \epsilon 
    ^2+\left(-\tfrac{1808 \zeta_3}{9}-\tfrac{123473}{324}\right) \epsilon 
    -550 \zeta_3 -\tfrac{14731}{72} \right] \\
     &{} \qquad  + r_c \left[ \left(\tfrac{128 \zeta_3}{9}+\tfrac{7495}{243}\right) \epsilon^2+\left(\tfrac{2504 \zeta_3}{9}+\tfrac{71765}{324}\right) \epsilon
     +396 \zeta_3+\tfrac{154 \epsilon ^3}{243}+\tfrac{30047}{72}\right] \\
     &{} \qquad  + r_c^2 \left[ \left(\tfrac{623}{27}-\tfrac{488 \zeta_3}{9}\right) \epsilon 
     ^2+\left(\tfrac{29753}{108}-\tfrac{5456 \zeta_3}{9}\right) \epsilon 
     -1694 \zeta_3+\tfrac{19613}{24} \right] \\
     &{} \qquad  + r_c^3 \left[ \tfrac{23 \epsilon }{4}+\tfrac{253}{8}\right] \big\} \,\alpha_g^3 \ + \ \\
     &{} \phantom{=\ } \big\{ \left[\left(36 \zeta_3+\tfrac{8017}{36}\right) \epsilon ^2+\left(396 \zeta_3+\tfrac{38797}{72}\right) \epsilon +1089 \zeta_3+\tfrac{379 \epsilon ^3}{18}-\tfrac{12947}{48} \right] \\
     &{} \qquad  + r_c \left[ \left(-54 \zeta_3-\tfrac{1184}{9}\right) \epsilon ^2+\left(-594 \zeta_3-\tfrac{45749}{72}\right) \epsilon -\tfrac{3267 \zeta_3}{2}-\tfrac{161 \epsilon ^3}{18}-\tfrac{24079}{24}\right] \\
     &{} \qquad + r_c^2 \left[\left(18 \zeta_3-\tfrac{3}{4}\right) \epsilon ^2+\left(198 \zeta_3-\tfrac{33}{4}\right) \epsilon +\tfrac{1089 \zeta_3}{2}-\tfrac{363}{16}\right]\big\} \, \alpha_g^2 \alpha_y \ + \ \\
     &{} \phantom{=\ } \big\{  \left[ \left(\tfrac{1659}{4}-12 \zeta_3\right) \epsilon ^2+(2475-132 \zeta_3) \epsilon -363 \zeta_3+23 \epsilon ^3+\tfrac{78287}{16}\right]\\
     &{} \qquad  + r_c  \left[ \left(12 \zeta_3+\tfrac{89}{4}\right) \epsilon ^2+(132 \zeta_3+154) \epsilon +363 \zeta_3+\epsilon ^3+\tfrac{5445}{16}\right]\big\}\,\alpha_g \alpha_y^2 \ + \ \\
     &{} \phantom{=\ } \big\{ \left[-\tfrac{11 \epsilon ^4}{3}-100 \epsilon ^3-986 \epsilon ^2-\tfrac{25267 \epsilon }{6}-\tfrac{105875}{16}\right] \\
     &{} \qquad  + r_c \left[\left(\tfrac{7}{3}-6 \zeta_3\right) \epsilon ^2+\left(\tfrac{77}{3}-66 \zeta_3\right) \epsilon -\tfrac{363 \zeta_3}{2}+\tfrac{847}{12}\right] \big\}\,\alpha_y^3 \ + \ \\
     &{} \phantom{=\ } \big\{ \left[-10 \epsilon ^3-165 \epsilon ^2-\tfrac{1815 \epsilon }{2}-\tfrac{6655}{4} \right] - r_c \left[55 + 10 \epsilon\right]\big\} \alpha_y^2 \alpha_u  -  
     r_c \left[ 20 \epsilon + 110 \right] \alpha_y^2 \alpha_v \ + \ \\
     &{} \phantom{=\ } \big\{ \left[12 \epsilon ^2+132 \epsilon +363\right] + 12 r_c \big\} \alpha_y \alpha_u^2 + 48 r_c \alpha_y \alpha_u \alpha_v + 12 r_c \left[1 + r_f\right] \alpha_y \alpha_v^2 \,.
  \end{aligned}
\end{equation}
Due to subleading corrections absent in the large-$N$ limit, the double-trace quartic  $\alpha_v$ makes a direct appearance in the gauge four loop $\beta$ function. 
The same happens for the three loop expressions of the Yukawa 
\begin{equation}
  \begin{aligned}
    \alpha_y^{-1}\, \beta_y^{(3)} &= \big\{ \left[-\tfrac{3 \epsilon ^3}{8}+\tfrac{59 \epsilon ^2}{16}+\tfrac{2595 \epsilon }{32}+\tfrac{17413}{64}\right] + r_c \left[(6 \zeta_3-28) \epsilon +39 \zeta_3-162\right] \big\} \, \alpha_y^3 \ + \ \\ 
    &{} \phantom{=\ } \, \big\{ \left[-19 \epsilon ^2-\tfrac{445 \epsilon }{2}-649\right] + r_c \left[19 \epsilon ^2+\tfrac{445 \epsilon }{2}+633\right] + 16 r_c^2\big\} \, \alpha_g \alpha_y^2 \ + \ \\
    &{} \phantom{=\ } \, \big\{ \left[\left(-36 \zeta_3-\tfrac{893}{8}\right) \epsilon -198 \zeta_3-17 \epsilon ^2-\tfrac{1217}{16}\right] + r_c \left[(54 \zeta_3+92) \epsilon +279 \zeta_3+17 \epsilon ^2+31\right]  \\
    &{} \qquad + r_c^2 \left[\left(\tfrac{157}{8}-18 \zeta_3\right) \epsilon -81 \zeta_3+\tfrac{721}{16}\right] \big\}\, \alpha_g^2 \alpha_y \ + \  \big\{ \left[\left(24 \zeta_3+\tfrac{649}{9}\right) \epsilon +132 \zeta_3+\tfrac{70 \epsilon ^2}{27}+\tfrac{641}{6}\right]\\
    &{} \qquad  + r_c \left[-\tfrac{70 \epsilon ^2}{27}-\tfrac{856 \epsilon }{9}-\tfrac{2413}{12}\right] + r_c^2\left[(23-24 \zeta_3) \epsilon -132 \zeta_3+62\right] + \tfrac{129}{4} r_c^3 \big\}\,\alpha_g^3 \ + \ \\
    &{} \phantom{=\ } \, \big\{ \left[12 \epsilon ^2+162 \epsilon +528\right] + 60 r_c + 30 ( \tfrac{11}{2} + \epsilon) r_f  \big\} \, \alpha_y^2 \alpha_u  + \big\{ 48 r_c + 60( \tfrac{11}{2} + \epsilon) r_f +  24 r_c r_f \big\}\, \alpha_y^2 \alpha_v \ + \ \\
    &{} \phantom{=\ } \, \big\{ \left[5 \epsilon +\tfrac{25}{2}\right] + r_f \left[85 \epsilon +\tfrac{905}{2}\right] \big\}\,\alpha_y \alpha_u^2 + \big\{r_f \left[100 \epsilon +490\right] + r_f^2 \left[80 \epsilon +440\right]\big\}\,\alpha_y \alpha_u \alpha_v \ + \ \\
    &{} \phantom{=\ } \, \big\{ r_f \left[5 \epsilon +\tfrac{25}{2}\right] + r_f^2 \left[85 \epsilon +\tfrac{905}{2}\right] \big\} \alpha_y \alpha_v^2 - \big\{ 8 + 32 r_f \big\} \alpha_u^3 - \big\{ 84 r_f + 36 r_f^2 \big\} \alpha_u^2 \alpha_v  \ - \ \\
    &{} \phantom{=\ } \, \big\{ 24 r_f + 96 r_f^2 \big\} \alpha_u \alpha_v^2 - \big\{ 4 r_f + 20 r_f^2 + 16 r_f^3\big\} \alpha_v^3 + \big\{ 4\left[ \tfrac{11}{2} + \epsilon \right] \left[1 - r_c \right]\left[1+ r_f\right]\big\}\, \alpha_g \alpha_y \alpha_u \ + \ \\
    &{} \phantom{=\ } \, \big\{  8\left[ \tfrac{11}{2} + \epsilon \right] \left[r_f - r_c r_f\right]\big\}\, \alpha_g \alpha_y \alpha_v \,,
  \end{aligned}
\end{equation}
as well as the single-trace coupling
\begin{equation}
  \begin{aligned}
    \beta_u^{(3)} &= \big\{ 104 + r_f \left[1152 \zeta_3+2360\right]\big\}\, \alpha_u^4 + \big\{ r_f \left[1536 \zeta_3+2912\right] + r_f^2 \left[6144 \zeta_3+6752\right]\big\} \, \alpha_u^3 \alpha_v  \ - \ \\
    &{} \phantom{=\ } \, \big\{ 280 r_f  - r_f^2 \left[9216 \zeta_3+12728\right]  \big\} \, \alpha_u^2 \alpha_v^2 - \big\{ 104 r_f  - r_f^2 \left[768 \zeta_3+1472\right]  - r_f^3 \left[5376 \zeta_3+6568\right]\big\} \, \alpha_u \alpha_v^3 \ + \ \\
    &{} \phantom{=\ } \,\big\{34 + 226 r_f\big\} \, \alpha_y \alpha_u^3 + 648 r_f \alpha_y \alpha_u^2 \alpha_v  + \big\{66 r_f + 642 r_f^2\big\} \,\alpha_y \alpha_u \alpha_v^2 \ + \ \\
    &{} \phantom{=\ } \,\big\{ \left[166 \epsilon +889\right] + r_f \left[(216 \zeta_3+156) \epsilon +1188 \zeta_3+858\right]\big\}\,\alpha_y^2 \alpha_u^2 \ + \ \\
    &{} \phantom{=\ } \,   \big\{r_f \left[(192 \zeta_3+734) \epsilon +1056 \zeta_3+3965\right]\big\}\,\alpha_y^2 \alpha_u \alpha_v \ + \ \\
    &{} \phantom{=\ } \, \big\{ r_f \left[64 \epsilon +352\right] + r_f^2 \left[(216 \zeta_3+136) \epsilon +1188 \zeta_3+748\right]\big\}\,\alpha_y^2 \alpha_v^2 \ + \ \\
    &{} \phantom{=\ } \, \big\{ \left[-\tfrac{315 \epsilon ^2}{4}-\tfrac{3209 \epsilon }{4}-\tfrac{32483}{16}\right] + r_c \left[12 \zeta_3-168\right]\big\} \alpha_y^3 \alpha_u - \big\{ r_c\left[152+ 96 \zeta_3\right] - 64\left[\tfrac{11}{2} + \epsilon\right] r_f \big\} \, \alpha_y^3 \alpha_v \ + \ \\
    &{} \phantom{=\ } \, \big\{ \left[\tfrac{13 \epsilon ^3}{4}+\tfrac{265 \epsilon ^2}{8}+\tfrac{1111 \epsilon }{16}-\tfrac{2541}{32}\right] + r_c \left[(20-24 \zeta_3) \epsilon -132 \zeta_3+110\right]\big\} \,\alpha_y^4 \ + \ \\
    &{} \phantom{=\ } \, \big\{ \left[(24 \zeta_3-5) \epsilon ^2+(264 \zeta_3-55) \epsilon +726 \zeta_3-\tfrac{605}{4}\right] + r_c \left[(5-24 \zeta_3) \epsilon ^2+(55-264 \zeta_3) \epsilon -726 \zeta_3+\tfrac{605}{4}\right]\big\} \, \alpha_g \alpha_y^3  \\
    &{} \phantom{=\ } \, + \big\{ \left[\left(\tfrac{149}{2}-120 \zeta_3\right) \epsilon -660 \zeta_3+\tfrac{1639}{4}\right] + r_c \left[\left(120 \zeta_3-\tfrac{149}{2}\right) \epsilon +660 \zeta_3-\tfrac{1639}{4}\right]\big\} \, \alpha_g \alpha_y^2 \alpha_u \ + \ \\
    &{} \phantom{=\ } \, \big\{r_f \left[(112-144 \zeta_3) \epsilon -792 \zeta_3+616\right] + r_c r_f \left[(144 \zeta_3-112) \epsilon +792 \zeta_3-616\right] \big\}\, \alpha_g \alpha_y^2 \alpha_v \ + \ \\
    &{} \phantom{=\ } \, \big\{ \left[96 \zeta_3-102\right] \left[1- r_c\right]\big\}\, \alpha_g \alpha_y \alpha_u^2 \ + \ \\
    &{} \phantom{=\ } \, \big\{r_f \left[288 \zeta_3-306\right] + r_f^2 \left[(306-288 \zeta_3) \epsilon ^2+(3366-3168 \zeta_3) \epsilon -8712 \zeta_3+\tfrac{18513}{2}\right]\big\}\,\alpha_g \alpha_y \alpha_u \alpha_v \ + \ \\
    &{} \phantom{=\ } \, \big\{ \left[5 \epsilon ^2+\tfrac{133 \epsilon }{4}+\tfrac{253}{8}\right] + r_c \left[(24 \zeta_3-66) \epsilon +132 \zeta_3-5 \epsilon ^2-\tfrac{847}{4}\right] + r_c^2 \left[\left(\tfrac{131}{4}-24 \zeta_3\right) \epsilon -132 \zeta_3+\tfrac{1441}{8}\right]\big\}\alpha_g^2 \alpha_y^2 \\
    &{} \phantom{=\ } \, + \big\{ \left[\tfrac{13}{4}-8 \epsilon\right] + r_c \left[-36 \zeta_3+8 \epsilon +\tfrac{53}{2}\right] + r_c^2 \left[36 \zeta_3-\tfrac{119}{4}\right]\big\}\,\alpha_g^2 \alpha_y \alpha_u\,.
  \end{aligned}
\end{equation}
Moreover, the three-loop $\beta$ function
\begin{equation}
  \begin{aligned}
    \beta_v^{(3)} &= \big\{ \left[384 \zeta_3+772\right] + r_f \left[1536 \zeta_3+1700\right] \big\}\,\alpha_u^4 + \big\{480 + r_f \left[4608 \zeta_3+9600\right]\big\}\,\alpha_u^3 \alpha_v \ + \ \\
    &{} \phantom{=\ } \, \big\{ 12 + r_f \left[1152 \zeta_3+6680\right] + r_f^2 \left[8064 \zeta_3+10476\right]\big\}\,\alpha_u^2 \alpha_v^2 + \big\{1264 r_f + r_f^2 \left[6144 \zeta_3+10544\right]\big\}\,\alpha_u \alpha_v^3 \ + \ \\
    &{} \phantom{=\ } \, \big\{132 r_f + r_f^2 \left[960 \zeta_3+1844\right] + r_f^3 \left[2112 \zeta_3+2960\right] \big\}\,\alpha_v^4 + 192\,\alpha_y \alpha_u^3 + \big\{66 + 642 r_f\big\}\,\alpha_y \alpha_u^2 \alpha_v \ + \ \\
    &{} \phantom{=\ } \, 648 r_f\,\alpha_y \alpha_u \alpha_v^2 + \big\{130 r_f + 322 r_f^2\big\} \,\alpha_y \alpha_v^3 + \left[(192 \zeta_3+187) \epsilon +1056 \zeta_3+\tfrac{1985}{2}\right] \alpha_y^2 \alpha_u^2 \ + \ \\
    &{} \phantom{=\ } \,\big\{ \left[(96 \zeta_3+152) \epsilon +528 \zeta_3+788\right] + r_f \left[(528 \zeta_3+132) \epsilon +2904 \zeta_3+726\right]\big\}\,\alpha_y^2 \alpha_u \alpha_v \ + \ \\
    &{} \phantom{=\ } \,\big\{ \left[41 \epsilon +\tfrac{427}{2}\right] + r_f \left[(192 \zeta_3+268) \epsilon +1056 \zeta_3+1426\right]\big\}\,\alpha_y^2  \alpha_v^2  \ + \ \\
    &{} \phantom{=\ } \, \left[(-96 \zeta_3-88) \epsilon ^2+(-1056 \zeta_3-904) \epsilon -2904 \zeta_3-2310\right] \alpha_y^3 \alpha_u \ + \ \\
    &{} \phantom{=\ } \, \big\{ \left[-\tfrac{187 \epsilon ^2}{4}-\tfrac{1801 \epsilon }{4}-\tfrac{16995}{16} \right] + r_c \left[12 \zeta_3-136\right]\big\}\, \alpha_y^3 \alpha_v + \left[-10 \epsilon ^3-183 \epsilon ^2-\tfrac{2211 \epsilon }{2}-\tfrac{8833}{4}\right] \alpha_y^4 \ + \ \\
    &{} \phantom{=\ } \, \big\{\left[(24 \zeta_3-2) \epsilon ^2+(264 \zeta_3-22) \epsilon +726 \zeta_3-\tfrac{121}{2}\right] + r_c \left[(2-24 \zeta_3) \epsilon ^2+(22-264 \zeta_3) \epsilon -726 \zeta_3+\tfrac{121}{2}\right]\big\}\, \alpha_g \alpha_y^3  \\
    &{} \phantom{=\ } \, +  \big\{ \left[(112-144 \zeta_3) \epsilon -792 \zeta_3+616\right]\left[1- r_c\right] \big\}\,\alpha_g \alpha_y^2 \alpha_u \ + \ \\
    &{} \phantom{=\ } \, \big\{ \left[\left(\tfrac{149}{2}-120 \zeta_3\right) \epsilon -660 \zeta_3+\tfrac{1639}{4}\right] \left[1 - r_c\right]\big\}\,\alpha_g \alpha_y^2 \alpha_v + \big\{ \left[144 \zeta_3-153\right] \left[1 - r_c\right]\big\}\,\alpha_g \alpha_y \alpha_u^2 \ + \ \\
    &{} \phantom{=\ } \, \big\{ \left[48 \zeta_3-51\right] + r_f \left[(51-48 \zeta_3) \epsilon ^2+(561-528 \zeta_3) \epsilon -1260 \zeta_3+\tfrac{5355}{4}\right]  \\
    &{} \qquad \, + r_f^2 \left[(204-192 \zeta_3) \epsilon ^2+(2244-2112 \zeta_3) \epsilon -5808 \zeta_3+6171\right]\big\}\,\alpha_g \alpha_y \alpha_v^2 \ + \  \\
    &{} \phantom{=\ } \, \big\{ \left[192 \zeta_3-204\right]\left[1-r_c\right]\big\}\,\alpha_g \alpha_y \alpha_u \alpha_v 
    + \big\{ \left[24 \epsilon^2+264 \epsilon +726 \right] \left[1 - r_c\right]\big\}\,\alpha_g^2 \alpha_y^2 \ + \ \\
    &{} \phantom{=\ } \, \big\{\left[\tfrac{13}{4}-8 \epsilon\right] + r_c \left[-36 \zeta_3+8 \epsilon +\tfrac{53}{2}\right] + r_c^2\left[36 \zeta_3-\tfrac{119}{4}\right]\big\}\,\alpha_g^2 \alpha_y \alpha_v 
  \end{aligned}
\end{equation}
ceases to be just quadratic in $\alpha_v$ due to subleading correction in $N$.

\section{Gauge-dependent Anomalous Dimensions}\label{sec:gauge-dep-ads}

In this appendix, we provide results for the gauge-dependent field strength anomalous dimensions of fermions $\psi$, gauge fields $A_\mu$ as well as their ghosts $c$, using the definition \eq{AD-definition}. 
We have computed these explicitly with $R_\xi$ gauge fixing, such that $\xi = 1$ corresponds to the 't Hooft -- Feynman gauge.
The scalar field anomalous dimension is gauge independent and provided in \eq{gamma_phi}.
The fermionic field anomalous dimension reads
\begin{equation}\label{eq:gamma_psi}
    \begin{aligned}
        \gamma_\psi^{(1)} &= \tfrac12\xi \alpha_g + \tfrac14(11 + 2 \epsilon)\alpha_y \,,\\[0.4em]
        \gamma_\psi^{(2)} &= \tfrac18 (  \xi^2 + 8 \xi - 4 \epsilon) \,\alpha_g^2 - \tfrac{1}{2}(11+2\epsilon)\alpha_y \alpha_g  
        - \tfrac1{32}(23 + 2 \epsilon)(11+2\epsilon)\alpha_y^2 \,,\\[0.4em]
        \gamma_\psi^{(3)} &= - \tfrac{11}8 (11 + 2 \epsilon) \alpha_u^2 \alpha_y 
        + \tfrac12 (11+2\epsilon)^2 \alpha_u \alpha_y^2 + \left(\tfrac{13387}{256} + \tfrac{2119}{128} \epsilon + \tfrac{49}{64} \epsilon^2 - \tfrac{3}{32} \epsilon^3  \right) \alpha_y^3 \\
        &\phantom{=}\ + \tfrac1{32} (11+2\epsilon)(137 + 48 \zeta_3 +24\epsilon) \alpha_y^2 \alpha_g  + \tfrac{1}{64}(11+2\epsilon)(77 - 192 \zeta_3 + 12 \epsilon) \alpha_y \alpha_g^2 \\
        &\phantom{=}\ + \big[ - \tfrac{331}{32} - \tfrac{21}{16} \zeta_3 - (\tfrac{111}{64} - \tfrac38 \zeta_3) \xi  + (\tfrac{39}{64} + \tfrac3{16}\zeta_3)\xi^2 + \tfrac5{32} \xi^3    - (\tfrac{109}{24} + \tfrac{17}{16} \xi) \epsilon + \tfrac5{18} \epsilon^2 \big] \alpha_g^3 
        \,.
    \end{aligned}
\end{equation}
The gauge field anomalous dimension is 
\begin{equation}\label{eq:gamma_gauge}
    \begin{aligned}
        \gamma_A^{(1)} &= \tfrac16 (9 + 3 \xi + 4 \epsilon) \alpha_g \,,\\[0.4em]
        \gamma_A^{(2)} &= \tfrac18 (  95 + 11 \xi + 2 \xi^2 + 28 \epsilon) \,\alpha_g^2  - \tfrac{1}{4}(11+2\epsilon)^2\alpha_y \alpha_g  \,,\\[0.4em]
        \gamma_A^{(3)} &= \tfrac18(11+2\epsilon)^2 (20+3\epsilon) \alpha_y^2 \alpha_g  -\tfrac{31}{32}(11+2\epsilon)^2 \alpha_g^2 \alpha_y \\
        &\phantom{=}\ + \big[
             \tfrac{2039}{96} - \tfrac{255}{16} \zeta_3 - (\tfrac9{32} - \tfrac34 \zeta_3)\xi   +
             (\tfrac{33}{32}+ \tfrac3{16} \zeta_3 )\xi^2 + \tfrac7{32}\xi^3 
             - (\tfrac{347}{72} + 3 \zeta_3 + \xi )\epsilon - \tfrac{49}{18}\epsilon^2
            \big] \alpha_g^3\,,
    \end{aligned}
\end{equation}
and the corresponding ghost has
\begin{equation}\label{eq:ghost}
    \begin{aligned}
        \gamma_c^{(1)} &= -\tfrac14 (3 - \xi) \alpha_g \,,\\[0.4em]
        \gamma_c^{(2)} &= \tfrac1{48} (  15 - 3 \xi + 20 \epsilon) \,\alpha_g^2 \,,  \\[0.4em]
        \gamma_c^{(3)} &= -\tfrac{23}{64} (11+2 \epsilon)^2 \alpha_g^2 \alpha_y  \\
        &\phantom{=}\  + \big[ \tfrac{3569}{192} + \tfrac{255}{32}\zeta_3 - \tfrac18 (15 + 3 \zeta_3) \xi + \tfrac{3}{32} (1 - \zeta_3) \xi^2 + \tfrac3{64}\xi^3  + (\tfrac{983}{144} + \tfrac32 \zeta_3 - \tfrac{7}{16}\xi ) \epsilon + \tfrac{35}{108} \epsilon^2
        \big]\alpha_g^3\,.
    \end{aligned}
\end{equation}
As the overall renormalisation of the gauge-fixing term cancels, the $\beta$ function of the gauge parameter reads
\begin{equation}\label{eq:dxi}
    \beta_\xi = - 2 \,\xi\,\gamma_A\,.
\end{equation}
We observe that \eqref{eq:dxi} has two types of fixed points due to either Landau gauge $(\xi^*=0)$, or  the vanishing of the gauge field anomalous dimension $(\gamma_A=0)$. The latter happens at
\begin{equation}\label{eq:xi*}
    \xi^* = -3 + 2.28\,\epsilon + 10.19\,\epsilon^2 + 21.92\,\epsilon^3 +  \mathcal{O}(\epsilon^4)\,.
\end{equation}
Moreover, we note that the critical exponent of the flow \eqref{eq:dxi} at the fixed point \eqref{eq:xi*}
\begin{equation}
    \frac{\partial \beta_\xi}{\partial \xi} \Big|_{\xi = \xi^*} = 1.368 \,\epsilon + 1.146\,\epsilon^2 + 13.83\,\epsilon^3+  \mathcal{O}(\epsilon^4)
\end{equation}
is manifestly positive. Hence, we conclude that the Landau gauge corresponds to an UV fixed point of the flow \eqref{eq:dxi}, whereas a vanishing gauge field anomalous dimension corresponds to an IR attractive fixed point.
\end{widetext}

\bibliographystyle{JHEP}
\bibliography{ref.bib}

\providecommand{\href}[2]{#2}\begingroup\raggedright\begin{thebibliography}{100}

\bibitem{Gross:1973id}
D.~J. Gross and F.~Wilczek, \emph{{Ultraviolet Behavior of Nonabelian Gauge
  Theories}}, \href{https://doi.org/10.1103/PhysRevLett.30.1343}{\emph{Phys.
  Rev. Lett.} {\bfseries 30} (1973) 1343}.

\bibitem{Politzer:1973fx}
H.~D. Politzer, \emph{{Reliable Perturbative Results for Strong
  Interactions?}},
  \href{https://doi.org/10.1103/PhysRevLett.30.1346}{\emph{Phys. Rev. Lett.}
  {\bfseries 30} (1973) 1346}.

\bibitem{Litim:2014uca}
D.~F. Litim and F.~Sannino, \emph{{Asymptotic safety guaranteed}},
  \href{https://doi.org/10.1007/JHEP12(2014)178}{\emph{JHEP} {\bfseries 12}
  (2014) 178} [\href{https://arxiv.org/abs/1406.2337}{{\ttfamily 1406.2337}}].

\bibitem{Bond:2016dvk}
A.~D. Bond and D.~F. Litim, \emph{{Theorems for Asymptotic Safety of Gauge
  Theories}}, \href{https://doi.org/10.1140/epjc/s10052-017-4976-5}{\emph{Eur.
  Phys. J. C} {\bfseries 77} (2017) 429}
  [\href{https://arxiv.org/abs/1608.00519}{{\ttfamily 1608.00519}}].

\bibitem{Bond:2018oco}
A.~D. Bond and D.~F. Litim, \emph{{Price of Asymptotic Safety}},
  \href{https://doi.org/10.1103/PhysRevLett.122.211601}{\emph{Phys. Rev. Lett.}
  {\bfseries 122} (2019) 211601}
  [\href{https://arxiv.org/abs/1801.08527}{{\ttfamily 1801.08527}}].

\bibitem{Bond:2017lnq}
A.~D. Bond and D.~F. Litim, \emph{{More asymptotic safety guaranteed}},
  \href{https://doi.org/10.1103/PhysRevD.97.085008}{\emph{Phys. Rev. D}
  {\bfseries 97} (2018) 085008}
  [\href{https://arxiv.org/abs/1707.04217}{{\ttfamily 1707.04217}}].

\bibitem{Bond:2017suy}
A.~D. Bond and D.~F. Litim, \emph{{Asymptotic safety guaranteed in
  supersymmetry}},
  \href{https://doi.org/10.1103/PhysRevLett.119.211601}{\emph{Phys. Rev. Lett.}
  {\bfseries 119} (2017) 211601}
  [\href{https://arxiv.org/abs/1709.06953}{{\ttfamily 1709.06953}}].

\bibitem{Bond:2017tbw}
A.~D. Bond, D.~F. Litim, G.~Medina~Vazquez and T.~Steudtner, \emph{{UV
  conformal window for asymptotic safety}},
  \href{https://doi.org/10.1103/PhysRevD.97.036019}{\emph{Phys. Rev. D}
  {\bfseries 97} (2018) 036019}
  [\href{https://arxiv.org/abs/1710.07615}{{\ttfamily 1710.07615}}].

\bibitem{Bond:2017sem}
A.~Bond and D.~F. Litim, \emph{{Interacting ultraviolet completions of
  four-dimensional gauge theories}},
  \href{https://doi.org/10.22323/1.256.0208}{\emph{PoS} {\bfseries LATTICE2016}
  (2017) 208}.

\bibitem{Buyukbese:2017ehm}
T.~Buyukbese and D.~F. Litim, \emph{{Asymptotic safety of gauge theories beyond
  marginal interactions}},
  \href{https://doi.org/10.22323/1.256.0233}{\emph{PoS} {\bfseries LATTICE2016}
  (2017) 233}.

\bibitem{Bond:2019npq}
A.~D. Bond, D.~F. Litim and T.~Steudtner, \emph{{Asymptotic safety with
  Majorana fermions and new large $N$ equivalences}},
  \href{https://doi.org/10.1103/PhysRevD.101.045006}{\emph{Phys. Rev. D}
  {\bfseries 101} (2020) 045006}
  [\href{https://arxiv.org/abs/1911.11168}{{\ttfamily 1911.11168}}].

\bibitem{Litim:2020jvl}
D.~F. Litim and T.~Steudtner, \emph{{ARGES \textendash{} Advanced
  Renormalisation Group Equation Simplifier}},
  \href{https://doi.org/10.1016/j.cpc.2021.108021}{\emph{Comput. Phys. Commun.}
  {\bfseries 265} (2021) 108021}
  [\href{https://arxiv.org/abs/2012.12955}{{\ttfamily 2012.12955}}].

\bibitem{Bond:2021tgu}
A.~D. Bond, D.~F. Litim and G.~M. Vazquez, \emph{{Conformal windows beyond
  asymptotic freedom}},
  \href{https://doi.org/10.1103/PhysRevD.104.105002}{\emph{Phys. Rev. D}
  {\bfseries 104} (2021) 105002}
  [\href{https://arxiv.org/abs/2107.13020}{{\ttfamily 2107.13020}}].

\bibitem{Bond:2022xvr}
A.~D. Bond and D.~F. Litim, \emph{{Asymptotic safety guaranteed for strongly
  coupled gauge theories}},
  \href{https://doi.org/10.1103/PhysRevD.105.105005}{\emph{Phys. Rev. D}
  {\bfseries 105} (2022) 105005}
  [\href{https://arxiv.org/abs/2202.08223}{{\ttfamily 2202.08223}}].

\bibitem{Bailin:1974bq}
D.~Bailin and A.~Love, \emph{{Asymptotic Near Freedom}},
  \href{https://doi.org/10.1016/0550-3213(74)90470-2}{\emph{Nucl. Phys. B}
  {\bfseries 75} (1974) 159}.

\bibitem{Weinberg:1980gg}
S.~Weinberg, \emph{{Ultraviolet Divergences in Quantum Theories of
  Gravitation}},  in \emph{General Relativity: An Einstein Centenary Survey},
  pp.~790--831, (1980).

\bibitem{Litim:2015iea}
D.~F. Litim, M.~Mojaza and F.~Sannino, \emph{{Vacuum stability of
  asymptotically safe gauge-Yukawa theories}},
  \href{https://doi.org/10.1007/JHEP01(2016)081}{\emph{JHEP} {\bfseries 01}
  (2016) 081} [\href{https://arxiv.org/abs/1501.03061}{{\ttfamily
  1501.03061}}].

\bibitem{Sannino:2014lxa}
F.~Sannino and I.~M. Shoemaker, \emph{{Asymptotically Safe Dark Matter}},
  \href{https://doi.org/10.1103/PhysRevD.92.043518}{\emph{Phys. Rev. D}
  {\bfseries 92} (2015) 043518}
  [\href{https://arxiv.org/abs/1412.8034}{{\ttfamily 1412.8034}}].

\bibitem{Nielsen:2015una}
N.~G. Nielsen, F.~Sannino and O.~Svendsen, \emph{{Inflation from Asymptotically
  Safe Theories}},
  \href{https://doi.org/10.1103/PhysRevD.91.103521}{\emph{Phys. Rev. D}
  {\bfseries 91} (2015) 103521}
  [\href{https://arxiv.org/abs/1503.00702}{{\ttfamily 1503.00702}}].

\bibitem{Rischke:2015mea}
D.~H. Rischke and F.~Sannino, \emph{{Thermodynamics of asymptotically safe
  theories}}, \href{https://doi.org/10.1103/PhysRevD.92.065014}{\emph{Phys.
  Rev. D} {\bfseries 92} (2015) 065014}
  [\href{https://arxiv.org/abs/1505.07828}{{\ttfamily 1505.07828}}].

\bibitem{Codello:2016muj}
A.~Codello, K.~Lang\ae{}ble, D.~F. Litim and F.~Sannino, \emph{{Conformal
  Gauge-Yukawa Theories away From Four Dimensions}},
  \href{https://doi.org/10.1007/JHEP07(2016)118}{\emph{JHEP} {\bfseries 07}
  (2016) 118} [\href{https://arxiv.org/abs/1603.03462}{{\ttfamily
  1603.03462}}].

\bibitem{Bond:2017wut}
A.~D. Bond, G.~Hiller, K.~Kowalska and D.~F. Litim, \emph{{Directions for model
  building from asymptotic safety}},
  \href{https://doi.org/10.1007/JHEP08(2017)004}{\emph{JHEP} {\bfseries 08}
  (2017) 004} [\href{https://arxiv.org/abs/1702.01727}{{\ttfamily
  1702.01727}}].

\bibitem{Dondi:2017civ}
N.~A. Dondi, V.~Prochazka and F.~Sannino, \emph{{Conformal Data of Fundamental
  Gauge-Yukawa Theories}},
  \href{https://doi.org/10.1103/PhysRevD.98.045002}{\emph{Phys. Rev. D}
  {\bfseries 98} (2018) 045002}
  [\href{https://arxiv.org/abs/1712.05388}{{\ttfamily 1712.05388}}].

\bibitem{Kowalska:2017fzw}
K.~Kowalska, A.~Bond, G.~Hiller and D.~Litim, \emph{{Towards an asymptotically
  safe completion of the Standard Model}},
  \href{https://doi.org/10.22323/1.314.0542}{\emph{PoS} {\bfseries EPS-HEP2017}
  (2017) 542}.

\bibitem{Abel:2017ujy}
S.~Abel and F.~Sannino, \emph{{Radiative symmetry breaking from interacting UV
  fixed points}}, \href{https://doi.org/10.1103/PhysRevD.96.056028}{\emph{Phys.
  Rev. D} {\bfseries 96} (2017) 056028}
  [\href{https://arxiv.org/abs/1704.00700}{{\ttfamily 1704.00700}}].

\bibitem{Christiansen:2017qca}
N.~Christiansen, A.~Eichhorn and A.~Held, \emph{{Is scale-invariance in
  gauge-Yukawa systems compatible with the graviton?}},
  \href{https://doi.org/10.1103/PhysRevD.96.084021}{\emph{Phys. Rev. D}
  {\bfseries 96} (2017) 084021}
  [\href{https://arxiv.org/abs/1705.01858}{{\ttfamily 1705.01858}}].

\bibitem{Sannino:2018suq}
F.~Sannino and V.~Skrinjar, \emph{{Instantons in asymptotically safe and free
  quantum field theories}},
  \href{https://doi.org/10.1103/PhysRevD.99.085010}{\emph{Phys. Rev. D}
  {\bfseries 99} (2019) 085010}
  [\href{https://arxiv.org/abs/1802.10372}{{\ttfamily 1802.10372}}].

\bibitem{Barducci:2018ysr}
D.~Barducci, M.~Fabbrichesi, C.~M. Nieto, R.~Percacci and V.~Skrinjar,
  \emph{{In search of a UV completion of the standard model \textemdash{}
  378,000 models that don\textquoteright{}t work}},
  \href{https://doi.org/10.1007/JHEP11(2018)057}{\emph{JHEP} {\bfseries 11}
  (2018) 057} [\href{https://arxiv.org/abs/1807.05584}{{\ttfamily
  1807.05584}}].

\bibitem{Hiller:2019mou}
G.~Hiller, C.~Hormigos-Feliu, D.~F. Litim and T.~Steudtner, \emph{{Anomalous
  magnetic moments from asymptotic safety}},
  \href{https://doi.org/10.1103/PhysRevD.102.071901}{\emph{Phys. Rev. D}
  {\bfseries 102} (2020) 071901}
  [\href{https://arxiv.org/abs/1910.14062}{{\ttfamily 1910.14062}}].

\bibitem{Hiller:2019tvg}
G.~Hiller, C.~Hormigos-Feliu, D.~F. Litim and T.~Steudtner,
  \emph{{Asymptotically safe extensions of the Standard Model with flavour
  phenomenology}},  in \emph{{54th Rencontres de Moriond on Electroweak
  Interactions and Unified Theories}}, pp.~415--418, 2019,
  \href{https://arxiv.org/abs/1905.11020}{{\ttfamily 1905.11020}}.

\bibitem{Hiller:2020fbu}
G.~Hiller, C.~Hormigos-Feliu, D.~F. Litim and T.~Steudtner, \emph{{Model
  Building from Asymptotic Safety with Higgs and Flavor Portals}},
  \href{https://doi.org/10.1103/PhysRevD.102.095023}{\emph{Phys. Rev. D}
  {\bfseries 102} (2020) 095023}
  [\href{https://arxiv.org/abs/2008.08606}{{\ttfamily 2008.08606}}].

\bibitem{Bissmann:2020lge}
S.~Bi\ss{}mann, G.~Hiller, C.~Hormigos-Feliu and D.~F. Litim,
  \emph{{Multi-lepton signatures of vector-like leptons with flavor}},
  \href{https://doi.org/10.1140/epjc/s10052-021-08886-3}{\emph{Eur. Phys. J. C}
  {\bfseries 81} (2021) 101}
  [\href{https://arxiv.org/abs/2011.12964}{{\ttfamily 2011.12964}}].

\bibitem{Bause:2021prv}
R.~Bause, G.~Hiller, T.~H\"ohne, D.~F. Litim and T.~Steudtner,
  \emph{{B-anomalies from flavorful U(1)$'$ extensions, safely}},
  \href{https://doi.org/10.1140/epjc/s10052-021-09957-1}{\emph{Eur. Phys. J. C}
  {\bfseries 82} (2022) 42} [\href{https://arxiv.org/abs/2109.06201}{{\ttfamily
  2109.06201}}].

\bibitem{Hiller:2022hgt}
G.~Hiller, D.~F. Litim and K.~Moch, \emph{{Fixed points in supersymmetric
  extensions of the standard model}},
  \href{https://doi.org/10.1140/epjc/s10052-022-10885-x}{\emph{Eur. Phys. J. C}
  {\bfseries 82} (2022) 952}
  [\href{https://arxiv.org/abs/2202.01264}{{\ttfamily 2202.01264}}].

\bibitem{Hiller:2022rla}
G.~Hiller, T.~H\"ohne, D.~F. Litim and T.~Steudtner, \emph{{Portals into Higgs
  vacuum stability}},
  \href{https://doi.org/10.1103/PhysRevD.106.115004}{\emph{Phys. Rev. D}
  {\bfseries 106} (2022) 115004}
  [\href{https://arxiv.org/abs/2207.07737}{{\ttfamily 2207.07737}}].

\bibitem{Hiller:2023bdb}
G.~Hiller, T.~H\"ohne, D.~F. Litim and T.~Steudtner, \emph{{Vacuum Stability as
  a Guide for Model Buislding}},  in \emph{{57th Rencontres de Moriond on
  Electroweak Interactions and Unified Theories}}, 5, 2023,
  \href{https://arxiv.org/abs/2305.18520}{{\ttfamily 2305.18520}}.

\bibitem{DelDebbio:2010zz}
L.~Del~Debbio, \emph{{The conformal window on the lattice}},
  \href{https://doi.org/10.22323/1.105.0004}{\emph{PoS} {\bfseries Lattice2010}
  (2014) 004} [\href{https://arxiv.org/abs/1102.4066}{{\ttfamily 1102.4066}}].

\bibitem{DiPietro:2020jne}
L.~Di~Pietro and M.~Serone, \emph{{Looking through the QCD Conformal Window
  with Perturbation Theory}},
  \href{https://doi.org/10.1007/JHEP07(2020)049}{\emph{JHEP} {\bfseries 07}
  (2020) 049} [\href{https://arxiv.org/abs/2003.01742}{{\ttfamily
  2003.01742}}].

\bibitem{Kluth:2022wgh}
Y.~Kluth, D.~F. Litim and M.~Reichert, \emph{{Spectral functions of gauge
  theories with Banks-Zaks fixed points}},
  \href{https://doi.org/10.1103/PhysRevD.107.025011}{\emph{Phys. Rev. D}
  {\bfseries 107} (2023) 025011}
  [\href{https://arxiv.org/abs/2207.14510}{{\ttfamily 2207.14510}}].

\bibitem{Bednyakov:2021qxa}
A.~Bednyakov and A.~Pikelner, \emph{{Four-Loop Gauge and Three-Loop Yukawa Beta
  Functions in a General Renormalizable Theory}},
  \href{https://doi.org/10.1103/PhysRevLett.127.041801}{\emph{Phys. Rev. Lett.}
  {\bfseries 127} (2021) 041801}
  [\href{https://arxiv.org/abs/2105.09918}{{\ttfamily 2105.09918}}].

\bibitem{Davies:2021mnc}
J.~Davies, F.~Herren and A.~E. Thomsen, \emph{{General gauge-Yukawa-quartic
  $\beta$-functions at 4-3-2-loop order}},
  \href{https://doi.org/10.1007/JHEP01(2022)051}{\emph{JHEP} {\bfseries 01}
  (2022) 051} [\href{https://arxiv.org/abs/2110.05496}{{\ttfamily
  2110.05496}}].

\bibitem{Veneziano:1976wm}
G.~Veneziano, \emph{{Some Aspects of a Unified Approach to Gauge, Dual and
  Gribov Theories}},
  \href{https://doi.org/10.1016/0550-3213(76)90412-0}{\emph{Nucl. Phys. B}
  {\bfseries 117} (1976) 519}.

\bibitem{tHooft:1973alw}
G.~'t~Hooft, \emph{{A Planar Diagram Theory for Strong Interactions}},
  \href{https://doi.org/10.1016/0550-3213(74)90154-0}{\emph{Nucl. Phys. B}
  {\bfseries 72} (1974) 461}.

\bibitem{Caswell:1974gg}
W.~E. Caswell, \emph{{Asymptotic Behavior of Nonabelian Gauge Theories to Two
  Loop Order}}, \href{https://doi.org/10.1103/PhysRevLett.33.244}{\emph{Phys.
  Rev. Lett.} {\bfseries 33} (1974) 244}.

\bibitem{Banks:1981nn}
T.~Banks and A.~Zaks, \emph{{On the Phase Structure of Vector-Like Gauge
  Theories with Massless Fermions}},
  \href{https://doi.org/10.1016/0550-3213(82)90035-9}{\emph{Nucl. Phys. B}
  {\bfseries 196} (1982) 189}.

\bibitem{Poole:2019kcm}
C.~Poole and A.~E. Thomsen, \emph{{Constraints on 3- and 4-loop
  $\beta$-functions in a general four-dimensional Quantum Field Theory}},
  \href{https://doi.org/10.1007/JHEP09(2019)055}{\emph{JHEP} {\bfseries 09}
  (2019) 055} [\href{https://arxiv.org/abs/1906.04625}{{\ttfamily
  1906.04625}}].

\bibitem{Valenti:2022uii}
A.~Valenti and L.~Vecchi, \emph{{Perturbative running of the topological
  angles}},  \href{https://arxiv.org/abs/2210.09328}{{\ttfamily 2210.09328}}.

\bibitem{Machacek:1984zw}
M.~E. Machacek and M.~T. Vaughn, \emph{{Two-loop renormalization group
  equations in a general quantum field theory: (III). Scalar quartic
  couplings}}, \href{https://doi.org/10.1016/0550-3213(85)90040-9}{\emph{Nucl.
  Phys.} {\bfseries B249} (1985) 70}.

\bibitem{Luo:2002ti}
M.~Luo, H.~Wang and Y.~Xiao, \emph{{Two-loop renormalization group equations in
  general gauge field theories}},
  \href{https://doi.org/10.1103/PhysRevD.67.065019}{\emph{Phys. Rev. D}
  {\bfseries 67} (2003) 065019}
  [\href{https://arxiv.org/abs/hep-ph/0211440}{{\ttfamily hep-ph/0211440}}].

\bibitem{Thomsen:2021ncy}
A.~E. Thomsen, \emph{{Introducing RGBeta: a Mathematica package for the
  evaluation of renormalization group $ \beta $-functions}},
  \href{https://doi.org/10.1140/epjc/s10052-021-09142-4}{\emph{Eur. Phys. J. C}
  {\bfseries 81} (2021) 408}
  [\href{https://arxiv.org/abs/2101.08265}{{\ttfamily 2101.08265}}].

\bibitem{Steudtner:FoRGEr}
T.~Steudtner, \emph{{FoRGEr}}, {\emph{{unpublished}} (2022) }.

\bibitem{Steudtner:2020tzo}
T.~Steudtner, \emph{{General scalar renormalisation group equations at
  three-loop order}},
  \href{https://doi.org/10.1007/JHEP12(2020)012}{\emph{JHEP} {\bfseries 12}
  (2020) 012} [\href{https://arxiv.org/abs/2007.06591}{{\ttfamily
  2007.06591}}].

\bibitem{Steudtner:2021fzs}
T.~Steudtner, \emph{{Towards general scalar-Yukawa renormalisation group
  equations at three-loop order}},
  \href{https://doi.org/10.1007/JHEP05(2021)060}{\emph{JHEP} {\bfseries 05}
  (2021) 060} [\href{https://arxiv.org/abs/2101.05823}{{\ttfamily
  2101.05823}}].

\bibitem{MaRTIn}
J.~Brod, E.~Stamou and T.~Steudtner, \emph{{MaRTIn}},
  {\emph{\textit{unplublished}} (2022) }.

\bibitem{Nogueira:1991ex}
P.~Nogueira, \emph{{Automatic Feynman graph generation}},
  \href{https://doi.org/10.1006/jcph.1993.1074}{\emph{J. Comput. Phys.}
  {\bfseries 105} (1993) 279}.

\bibitem{Kuipers:2012rf}
J.~Kuipers, T.~Ueda, J.~A.~M. Vermaseren and J.~Vollinga, \emph{{FORM version
  4.0}}, \href{https://doi.org/10.1016/j.cpc.2012.12.028}{\emph{Comput. Phys.
  Commun.} {\bfseries 184} (2013) 1453}
  [\href{https://arxiv.org/abs/1203.6543}{{\ttfamily 1203.6543}}].

\bibitem{Misiak:1994zw}
M.~Misiak and M.~Munz, \emph{{Two loop mixing of dimension five flavor changing
  operators}}, \href{https://doi.org/10.1016/0370-2693(94)01553-O}{\emph{Phys.
  Lett. B} {\bfseries 344} (1995) 308}
  [\href{https://arxiv.org/abs/hep-ph/9409454}{{\ttfamily hep-ph/9409454}}].

\bibitem{Chetyrkin:1997fm}
K.~G. Chetyrkin, M.~Misiak and M.~Munz, \emph{{Beta functions and anomalous
  dimensions up to three loops}},
  \href{https://doi.org/10.1016/S0550-3213(98)00122-9}{\emph{Nucl. Phys. B}
  {\bfseries 518} (1998) 473}
  [\href{https://arxiv.org/abs/hep-ph/9711266}{{\ttfamily hep-ph/9711266}}].

\bibitem{Chetyrkin:2012rz}
K.~Chetyrkin and M.~Zoller, \emph{{Three-loop $\beta$-functions for top-Yukawa
  and the Higgs self-interaction in the standard model}},
  \href{https://doi.org/10.1007/JHEP06(2012)033}{\emph{JHEP} {\bfseries 06}
  (2012) 033} [\href{https://arxiv.org/abs/1205.2892}{{\ttfamily 1205.2892}}].

\bibitem{Lee:2012cn}
R.~N. Lee, \emph{{Presenting LiteRed: a tool for the Loop InTEgrals
  REDuction}},  \href{https://arxiv.org/abs/1212.2685}{{\ttfamily 1212.2685}}.

\bibitem{Lee:2013mka}
R.~N. Lee, \emph{{LiteRed 1.4: a powerful tool for reduction of multiloop
  integrals}}, \href{https://doi.org/10.1088/1742-6596/523/1/012059}{\emph{J.
  Phys. Conf. Ser.} {\bfseries 523} (2014) 012059}
  [\href{https://arxiv.org/abs/1310.1145}{{\ttfamily 1310.1145}}].

\bibitem{Bobeth:1999mk}
C.~Bobeth, M.~Misiak and J.~Urban, \emph{{Photonic penguins at two loops and
  $m_t$ dependence of $BR[B \to X_s l^+ l^-]$}},
  \href{https://doi.org/10.1016/S0550-3213(00)00007-9}{\emph{Nucl. Phys. B}
  {\bfseries 574} (2000) 291}
  [\href{https://arxiv.org/abs/hep-ph/9910220}{{\ttfamily hep-ph/9910220}}].

\bibitem{Martin:2016bgz}
S.~P. Martin and D.~G. Robertson, \emph{{Evaluation of the general 3-loop
  vacuum Feynman integral}},
  \href{https://doi.org/10.1103/PhysRevD.95.016008}{\emph{Phys. Rev. D}
  {\bfseries 95} (2017) 016008}
  [\href{https://arxiv.org/abs/1610.07720}{{\ttfamily 1610.07720}}].

\bibitem{Jegerlehner:2000dz}
F.~Jegerlehner, \emph{{Facts of life with $\gamma_5$}},
  \href{https://doi.org/10.1007/s100520100573}{\emph{Eur. Phys. J. C}
  {\bfseries 18} (2001) 673}
  [\href{https://arxiv.org/abs/hep-th/0005255}{{\ttfamily hep-th/0005255}}].

\bibitem{Poole:2019txl}
C.~Poole and A.~Thomsen, \emph{{Weyl Consistency Conditions and $\gamma_5$}},
  \href{https://doi.org/10.1103/PhysRevLett.123.041602}{\emph{Phys. Rev. Lett.}
  {\bfseries 123} (2019) 041602}
  [\href{https://arxiv.org/abs/1901.02749}{{\ttfamily 1901.02749}}].

\bibitem{Davies:2019onf}
J.~Davies, F.~Herren, C.~Poole, M.~Steinhauser and A.~E. Thomsen, \emph{{Gauge
  Coupling $\beta$ Functions to Four-Loop Order in the Standard Model}},
  \href{https://doi.org/10.1103/PhysRevLett.124.071803}{\emph{Phys. Rev. Lett.}
  {\bfseries 124} (2020) 071803}
  [\href{https://arxiv.org/abs/1912.07624}{{\ttfamily 1912.07624}}].

\bibitem{Bednyakov:2012en}
A.~Bednyakov, A.~Pikelner and V.~Velizhanin, \emph{{Yukawa coupling
  beta-functions in the Standard Model at three loops}},
  \href{https://doi.org/10.1016/j.physletb.2013.04.038}{\emph{Phys. Lett. B}
  {\bfseries 722} (2013) 336}
  [\href{https://arxiv.org/abs/1212.6829}{{\ttfamily 1212.6829}}].

\bibitem{Bednyakov:2014pia}
A.~Bednyakov, A.~Pikelner and V.~Velizhanin, \emph{{Three-loop SM
  beta-functions for matrix Yukawa couplings}},
  \href{https://doi.org/10.1016/j.physletb.2014.08.049}{\emph{Phys. Lett. B}
  {\bfseries 737} (2014) 129}
  [\href{https://arxiv.org/abs/1406.7171}{{\ttfamily 1406.7171}}].

\bibitem{Herren:2017uxn}
F.~Herren, L.~Mihaila and M.~Steinhauser, \emph{{Gauge and Yukawa coupling beta
  functions of two-Higgs-doublet models to three-loop order}},
  \href{https://doi.org/10.1103/PhysRevD.97.015016}{\emph{Phys. Rev. D}
  {\bfseries 97} (2018) 015016}
  [\href{https://arxiv.org/abs/1712.06614}{{\ttfamily 1712.06614}}].

\bibitem{Jack:2016tpp}
I.~Jack and H.~Osborn, \emph{{Scheme Dependence and Multiple Couplings}},
  \href{https://arxiv.org/abs/1606.02571}{{\ttfamily 1606.02571}}.

\bibitem{Herren:2021yur}
F.~Herren and A.~E. Thomsen, \emph{{On ambiguities and divergences in
  perturbative renormalization group functions}},
  \href{https://doi.org/10.1007/JHEP06(2021)116}{\emph{JHEP} {\bfseries 06}
  (2021) 116} [\href{https://arxiv.org/abs/2104.07037}{{\ttfamily
  2104.07037}}].

\bibitem{Machacek:1983tz}
M.~E. Machacek and M.~T. Vaughn, \emph{{Two-loop renormalization group
  equations in a general quantum field theory: (I). Wave function
  renormalization}},
  \href{https://doi.org/10.1016/0550-3213(83)90610-7}{\emph{Nucl. Phys.}
  {\bfseries B222} (1983) 83}.

\bibitem{Machacek:1983fi}
M.~E. Machacek and M.~T. Vaughn, \emph{{Two-loop renormalization group
  equations in a general quantum field theory (II). Yukawa couplings}},
  \href{https://doi.org/10.1016/0550-3213(84)90533-9}{\emph{Nucl. Phys.}
  {\bfseries B236} (1984) 221}.

\bibitem{Pickering:2001aq}
A.~Pickering, J.~Gracey and D.~Jones, \emph{{Three loop gauge $\beta$-function
  for the most general single gauge-coupling theory}},
  \href{https://doi.org/10.1016/S0370-2693(01)00624-4}{\emph{Phys. Lett. B}
  {\bfseries 510} (2001) 347}
  [\href{https://arxiv.org/abs/hep-ph/0104247}{{\ttfamily hep-ph/0104247}}].

\bibitem{Mihaila:2014caa}
L.~Mihaila, \emph{{Three-loop gauge beta function in non-simple gauge groups}},
  \href{https://doi.org/10.22323/1.197.0060}{\emph{PoS} {\bfseries RADCOR2013}
  (2013) 060}.

\bibitem{Schienbein:2018fsw}
I.~Schienbein, F.~Staub, T.~Steudtner and K.~Svirina, \emph{{Revisiting RGEs
  for general gauge theories}},
  \href{https://doi.org/10.1016/j.nuclphysb.2018.12.001}{\emph{Nucl. Phys. B}
  {\bfseries 939} (2019) 1} [\href{https://arxiv.org/abs/1809.06797}{{\ttfamily
  1809.06797}}].

\bibitem{Bednyakov:2013eba}
A.~Bednyakov, A.~Pikelner and V.~Velizhanin, \emph{{Higgs self-coupling
  beta-function in the Standard Model at three loops}},
  \href{https://doi.org/10.1016/j.nuclphysb.2013.07.015}{\emph{Nucl. Phys. B}
  {\bfseries 875} (2013) 552}
  [\href{https://arxiv.org/abs/1303.4364}{{\ttfamily 1303.4364}}].

\bibitem{Chetyrkin:2013wya}
K.~Chetyrkin and M.~Zoller, \emph{{$\beta$-function for the Higgs
  self-interaction in the Standard Model at three-loop level}},
  \href{https://doi.org/10.1007/JHEP04(2013)091}{\emph{JHEP} {\bfseries 04}
  (2013) 091} [\href{https://arxiv.org/abs/1303.2890}{{\ttfamily 1303.2890}}].

\bibitem{Bednyakov:2013cpa}
A.~Bednyakov, A.~Pikelner and V.~Velizhanin, \emph{{Three-loop Higgs
  self-coupling beta-function in the Standard Model with complex Yukawa
  matrices}},
  \href{https://doi.org/10.1016/j.nuclphysb.2013.12.012}{\emph{Nucl. Phys. B}
  {\bfseries 879} (2014) 256}
  [\href{https://arxiv.org/abs/1310.3806}{{\ttfamily 1310.3806}}].

\bibitem{Zerf:2018csr}
N.~Zerf, P.~Marquard, R.~Boyack and J.~Maciejko, \emph{{Critical behavior of
  the QED$_3$-Gross-Neveu-Yukawa model at four loops}},
  \href{https://doi.org/10.1103/PhysRevB.98.165125}{\emph{Phys. Rev. B}
  {\bfseries 98} (2018) 165125}
  [\href{https://arxiv.org/abs/1808.00549}{{\ttfamily 1808.00549}}].

\bibitem{Gracey:1996he}
J.~A. Gracey, \emph{{The QCD Beta function at O(1/N(f))}},
  \href{https://doi.org/10.1016/0370-2693(96)00105-0}{\emph{Phys. Lett. B}
  {\bfseries 373} (1996) 178}
  [\href{https://arxiv.org/abs/hep-ph/9602214}{{\ttfamily hep-ph/9602214}}].

\bibitem{Luthe:2016ima}
T.~Luthe, A.~Maier, P.~Marquard and Y.~Schr\"oder, \emph{{Towards the five-loop
  Beta function for a general gauge group}},
  \href{https://doi.org/10.1007/JHEP07(2016)127}{\emph{JHEP} {\bfseries 07}
  (2016) 127} [\href{https://arxiv.org/abs/1606.08662}{{\ttfamily
  1606.08662}}].

\bibitem{Baikov:2016tgj}
P.~A. Baikov, K.~G. Chetyrkin and J.~H. K\"uhn, \emph{{Five-Loop Running of the
  QCD coupling constant}},
  \href{https://doi.org/10.1103/PhysRevLett.118.082002}{\emph{Phys. Rev. Lett.}
  {\bfseries 118} (2017) 082002}
  [\href{https://arxiv.org/abs/1606.08659}{{\ttfamily 1606.08659}}].

\bibitem{Herzog:2017ohr}
F.~Herzog, B.~Ruijl, T.~Ueda, J.~Vermaseren and A.~Vogt, \emph{The five-loop
  beta function of yang-mills theory with fermions},
  \href{https://doi.org/10.1007/JHEP02(2017)090}{\emph{JHEP} {\bfseries 02}
  (2017) 090} [\href{https://arxiv.org/abs/1701.01404}{{\ttfamily
  1701.01404}}].

\bibitem{Luthe:2017ttg}
T.~Luthe, A.~Maier, P.~Marquard and Y.~Schröder, \emph{The five-loop beta
  function for a general gauge group and anomalous dimensions beyond feynman
  gauge}, \href{https://doi.org/10.1007/JHEP10(2017)166}{\emph{JHEP} {\bfseries
  10} (2017) 166} [\href{https://arxiv.org/abs/1709.07718}{{\ttfamily
  1709.07718}}].

\bibitem{Chetyrkin:2017bjc}
K.~G. Chetyrkin, G.~Falcioni, F.~Herzog and J.~A.~M. Vermaseren,
  \emph{{Five-loop renormalisation of QCD in covariant gauges}},
  \href{https://doi.org/10.1007/JHEP10(2017)179}{\emph{JHEP} {\bfseries 10}
  (2017) 179} [\href{https://arxiv.org/abs/1709.08541}{{\ttfamily
  1709.08541}}].

\bibitem{Bednyakov:2021ojn}
A.~Bednyakov and A.~Pikelner, \emph{{Six-loop beta functions in general scalar
  theory}}, \href{https://doi.org/10.1007/JHEP04(2021)233}{\emph{JHEP}
  {\bfseries 04} (2021) 233}
  [\href{https://arxiv.org/abs/2102.12832}{{\ttfamily 2102.12832}}].

\bibitem{Steudtner:2020jcj}
T.~Steudtner, \emph{{Asymptotic safety: from perturbatively exact models to
  particle physics}}, Ph.D. thesis, Sussex U., 2020.

\bibitem{Laporta:2002pg}
S.~Laporta, \emph{{High precision epsilon expansions of massive four loop
  vacuum bubbles}},
  \href{https://doi.org/10.1016/S0370-2693(02)02910-6}{\emph{Phys. Lett. B}
  {\bfseries 549} (2002) 115}
  [\href{https://arxiv.org/abs/hep-ph/0210336}{{\ttfamily hep-ph/0210336}}].

\bibitem{Schroder:2002re}
Y.~Schroder, \emph{{Automatic reduction of four loop bubbles}},
  \href{https://doi.org/10.1016/S0920-5632(03)80208-6}{\emph{Nucl. Phys. B
  Proc. Suppl.} {\bfseries 116} (2003) 402}
  [\href{https://arxiv.org/abs/hep-ph/0211288}{{\ttfamily hep-ph/0211288}}].

\bibitem{Czakon:2004bu}
M.~Czakon, \emph{The four-loop qcd $beta$-function and anomalous dimensions},
  \href{https://doi.org/10.1016/j.nuclphysb.2005.01.012}{\emph{Nucl.Phys.B}
  {\bfseries 710} (2005) 485}
  [\href{https://arxiv.org/abs/hep-ph/0411261}{{\ttfamily hep-ph/0411261}}].

\bibitem{Schroder:2005va}
Y.~Schroder and A.~Vuorinen, \emph{{High-precision epsilon expansions of
  single-mass-scale four-loop vacuum bubbles}},
  \href{https://doi.org/10.1088/1126-6708/2005/06/051}{\emph{JHEP} {\bfseries
  06} (2005) 051} [\href{https://arxiv.org/abs/hep-ph/0503209}{{\ttfamily
  hep-ph/0503209}}].

\bibitem{Luthe:2015ngq}
T.~Luthe, \emph{{Fully massive vacuum integrals at 5 loops}}, Ph.D. thesis,
  Bielefeld U., 2015.

\bibitem{Pikelner:2017tgv}
A.~Pikelner, \emph{{FMFT: Fully Massive Four-loop Tadpoles}},
  \href{https://doi.org/10.1016/j.cpc.2017.11.017}{\emph{Comput. Phys. Commun.}
  {\bfseries 224} (2018) 282}
  [\href{https://arxiv.org/abs/1707.01710}{{\ttfamily 1707.01710}}].

\bibitem{Pomoni:2009joh}
E.~Pomoni and L.~Rastelli, \emph{{Large N Field Theory and AdS Tachyons}},
  \href{https://doi.org/10.1088/1126-6708/2009/04/020}{\emph{JHEP} {\bfseries
  04} (2009) 020} [\href{https://arxiv.org/abs/0805.2261}{{\ttfamily
  0805.2261}}].

\bibitem{Martin:1993zk}
S.~P. Martin and M.~T. Vaughn, \emph{{Two-loop renormalization group equations
  for soft supersymmetry-breaking couplings}},
  \href{https://doi.org/10.1103/PhysRevD.50.2282}{\emph{Phys. Rev. D}
  {\bfseries 50} (1994) 2282}
  [\href{https://arxiv.org/abs/hep-ph/9311340}{{\ttfamily hep-ph/9311340}}].

\bibitem{Paterson:1980fc}
A.~Paterson, \emph{{{Coleman-Weinberg} symmetry breaking in the chiral $SU(N)
  \times SU(N)$ linear $\sigma$ model}},
  \href{https://doi.org/10.1016/0550-3213(81)90489-2}{\emph{Nucl. Phys. B}
  {\bfseries 190} (1981) 188}.

\bibitem{Benini:2019dfy}
F.~Benini, C.~Iossa and M.~Serone, \emph{{Conformality Loss, Walking, and 4D
  Complex Conformal Field Theories at Weak Coupling}},
  \href{https://doi.org/10.1103/PhysRevLett.124.051602}{\emph{Phys. Rev. Lett.}
  {\bfseries 124} (2020) 051602}
  [\href{https://arxiv.org/abs/1908.04325}{{\ttfamily 1908.04325}}].

\bibitem{Weinberg:1978kz}
S.~Weinberg, \emph{{Phenomenological Lagrangians}},
  \href{https://doi.org/10.1016/0378-4371(79)90223-1}{\emph{Physica A}
  {\bfseries 96} (1979) 327}.

\bibitem{CW2023}
D.~F. Litim, N.~Riyaz, E.~Stamou and T.~Steudtner, \emph{\textit{in
  preparation}}, .

\bibitem{Luty:2012ww}
M.~A. Luty, J.~Polchinski and R.~Rattazzi, \emph{{The $a$-theorem and the
  Asymptotics of 4D Quantum Field Theory}},
  \href{https://doi.org/10.1007/JHEP01(2013)152}{\emph{JHEP} {\bfseries 01}
  (2013) 152} [\href{https://arxiv.org/abs/1204.5221}{{\ttfamily 1204.5221}}].

\bibitem{Mack:1975je}
G.~Mack, \emph{{All unitary ray representations of the conformal group SU(2,2)
  with positive energy}},
  \href{https://doi.org/10.1007/BF01613145}{\emph{Commun. Math. Phys.}
  {\bfseries 55} (1977) 1}.

\bibitem{Aarts:2023vsf}
G.~Aarts et~al., \emph{{Phase Transitions in Particle Physics - Results and
  Perspectives from Lattice Quantum Chromo-Dynamics}},  in \emph{{Phase
  Transitions in Particle Physics}: {Results and Perspectives from Lattice
  Quantum Chromo-Dynamics}}, 1, 2023,
  \href{https://arxiv.org/abs/2301.04382}{{\ttfamily 2301.04382}}.

\bibitem{Cardy:1996xt}
J.~L. Cardy, \emph{{Scaling and renormalization in statistical physics}}.
  Cambridge University Press, 1996.

\bibitem{Codello:2017hhh}
A.~Codello, M.~Safari, G.~P. Vacca and O.~Zanusso, \emph{{Functional
  perturbative RG and CFT data in the $\epsilon$-expansion}},
  \href{https://doi.org/10.1140/epjc/s10052-017-5505-2}{\emph{Eur. Phys. J. C}
  {\bfseries 78} (2018) 30} [\href{https://arxiv.org/abs/1705.05558}{{\ttfamily
  1705.05558}}].

\bibitem{Li:2020bnb}
Z.~Li and D.~Poland, \emph{{Searching for gauge theories with the conformal
  bootstrap}}, \href{https://doi.org/10.1007/JHEP03(2021)172}{\emph{JHEP}
  {\bfseries 03} (2021) 172}
  [\href{https://arxiv.org/abs/2005.01721}{{\ttfamily 2005.01721}}].

\end{thebibliography}\endgroup
\end{document}